\documentclass{cernyrep}
\usepackage[utf8]{inputenc}
\usepackage[T1]{fontenc}
\usepackage{inconsolata}
\usepackage[frozencache,cachedir=_minted-main]{minted}
\usepackage[capitalise]{cleveref}
\usepackage{hyperref}
\hypersetup{
  linkcolor  = black, % tab of contents and eq/tab/fig links in text
  citecolor  = blue,
  urlcolor   = blue,
  colorlinks = true,
  breaklinks = true
}
\usepackage{dirtree}
\usepackage{makecell}

\graphicspath{{./Figures/}{./}}

\makeatletter
\AtBeginEnvironment{minted}{%
  \let\bfseries\mdseries
  \let\mintedorigfcolorbox\fcolorbox
  \renewcommand{\fcolorbox}[4][]{#4}%
}
\AtEndEnvironment{minted}{%
  \let\fcolorbox\mintedorigfcolorbox
}
\makeatother

\providecommand{\tightlist}{%
}

\newenvironment{SpecSection}{%
  \begingroup
  \let\origsubsubsection\subsubsection
  \renewcommand{\subsubsection}[1]{%
    \origsubsubsection*{\emph{##1}}% toggles italic → upright if already italic
  }%
}{%
  \endgroup
}

\begin{document}

\preprint{CERN-LHCEFTWG-2025-001\\ CERN-LPCC-2025-003\\LHCHWG-2025-050\\MITP-25-066}
%\date{October, 2025}

\title{
{\normalsize \textmd{Joint LHC EFT Working Group \& LHC Higgs Working Group Note:}}\\
{\large \texttt{POPxf}: An Exchange Format for Polynomial Observable Predictions}
}

%%%%%%%%%%%% authors
\author{
Ilaria Brivio\aff{A} (ed.),
Ken Mimasu\aff{B} (ed.),
Peter Stangl\aff{C} (ed.),
\\
Anke Biek\"otter\aff{KIT},
Ana R. Cueto Gómez\aff{UAM},
Charlotte Knight\aff{IC},
Luca Mantani\aff{IFIC},
Eleonora Rossi\aff{OX},
Alejo N. Rossia\twoaff{UNIPD}{INFNPD},
Aleks Smolkovi\v{c}\aff{IJS}
}

%%%%%%%%%%%% affiliations
\institute{
\naff{A}{Dipartimento di Fisica e Astronomia, Universit\`a di Bologna and INFN Bologna, Italy}
\naff{B}{School of Physics and Astronomy, University of Southampton, Highfield,\newline Southampton~SO17~1BJ, United Kingdom}
\naff{C}{Institute of Physics, Johannes Gutenberg University Mainz, Staudingerweg 7, 55128 Mainz, Germany}
\naff{KIT}{Karlsruhe Institute of Technology, Institute for Theoretical Particle Physics, Wolfgang-Gaede-Straße 1, 76131
Karlsruhe, Germany}
\naff{UAM}{Universidad Autónoma de Madrid, Campus Universitario de Cantoblanco, Madrid, Spain}
\naff{IC}{Imperial College, London, United Kingdom}
\naff{IFIC}{Instituto de Fisica Corpuscular (IFIC), Universidad de Valencia-CSIC, E-46980 Valencia, Spain}
\naff{OX}{University of Oxford, Oxford, United Kingdom}
\naff{UNIPD}{Dipartimento di Fisica e Astronomia “G. Galilei”, Universit\`a di Padova, Via F. Marzolo 8, I-35131, Padova, Italy}
\naff{INFNPD}{Istituto Nazionale di Fisica Nucleare, Sezione di Padova, Via F. Marzolo 8, I-35131, Padova, Italy}
\naff{IJS}{Jo\v{z}ef Stefan Institute, Jamova 39, 1000 Ljubljana, Slovenia}
}

\begin{abstract}
We introduce the Polynomial Observable Prediction Exchange Format, \texttt{POPxf}, a structured, machine-readable data format for the publication and exchange of semi-analytical theoretical predictions in high energy physics. The format is designed to encode observables that can be expressed in terms of polynomials in model parameters, with particular emphasis on Effective Field Theory applications. All relevant assumptions and metadata are recorded explicitly, and the treatment of uncertainties and correlations is flexible enough to capture parameter-dependent effects. The format aims to improve reproducibility, facilitate global fits and reinterpretations, and streamline the use of theoretical predictions across the particle physics community.
\end{abstract}


\addtocounter{table}{-1}% the affiliations longtable in \maketitle steps the table counter
\maketitle

\tableofcontents

%%%%%%%%%%%%%%%%%%%%%%%%%%%%%%%%%%%%%%%%%%%%%%%
\section{Introduction}\label{sec:intro}
Theoretical predictions play a central role in particle physics phenomenology, allowing the interpretation of experimental data in terms of the underlying model parameters. In many cases, such predictions can be expressed in terms of polynomials in these parameters. A prominent example is observables in an Effective Field Theory (EFT), where they depend on a set of Wilson coefficients. In particular, the amplitudes in an EFT are often linear in the Wilson coefficients, such that the associated observables, depending on squared amplitudes, are given in terms of quadratic polynomials in the Wilson coefficients. This structure arises generically in analyses within the Standard Model Effective Field Theory (SMEFT)~\cite{Buchmuller:1985jz,Grzadkowski:2010es} and the Weak Effective Theory (WET)~\cite{Buchalla:1995vs,Jenkins:2017jig} truncated at dimension six. Other examples beyond EFT include observables in flavour physics that depend on hadronic form factors, which enter amplitudes linearly and are themselves typically expressed using polynomial parameterisations.

Despite the ubiquity of such polynomial expressions, they are rarely published in their entirety and often remain hidden in various public and private codebases. Even when predictions based on polynomial expressions are made accessible, this is often done using varying conventions, with insufficient metadata, and/or in formats that are not machine-readable. Such practices complicate reproducibility, hinder validation and cross-comparison, and lead to significant duplication of effort across the community. A standardised, machine-readable data format for sharing such predictions would address many of these issues. It would facilitate the reuse of theoretical predictions in global fits, reinterpretations, and future measurements. It would also help document and preserve the assumptions -- such as EFT basis choices, renormalisation scales, input parameter values, or normalisation conventions -- that are crucial for the correct use and interpretation of the results.

This note proposes the Polynomial Observable Prediction Exchange Format, \texttt{POPxf}, a structured data format designed to encode polynomial parameterisations of observables as functions of model parameters, with a focus on EFT applications, while remaining general enough to support other use cases. The format supports arbitrary polynomial degree and is sufficiently general to describe both single-valued and multi-valued observables (such as binned distributions), groups of related observables, as well as observables defined through functions of multiple polynomial components (such as branching fractions).

Returning to the example of EFT predictions, where observables depend on a set of beyond-the-standard-model (BSM) Wilson coefficients, $C_i$, the scattering amplitude for a given process often takes the form
\begin{equation}
\mathcal{A} = \mathcal{A}_{\mathrm{SM}} + \sum_i C_i\, \mathcal{A}_i,
\end{equation}
so that the corresponding observable $O$, typically proportional to $|\mathcal{A}|^2$, becomes a quadratic polynomial in the $C_i$:
\begin{equation}
O = O_{\mathrm{SM}} + \sum_i C_i\, O_i^{\mathrm{int}} + \sum_{i \le j} C_i C_j\, O_{ij}^{\mathrm{quad}}.
\end{equation}
Here, $O_{\mathrm{SM}}$ is the Standard Model (SM) prediction, $O_i^{\mathrm{int}}$ encodes the interference between SM and BSM amplitudes, and $O_{ij}^{\mathrm{quad}}$ captures the pure BSM-squared contributions. This structure is general for dimension-six SMEFT and WET analyses, for which the expansion of the amplitude is truncated at $\mathcal{O}(1/\Lambda^2)$, where $\Lambda$ is the cutoff parameter of the EFT.
While quadratic polynomial structures are common, higher-order terms can arise in EFTs when including dimension-eight operators or computing beyond leading order. The proposed data format supports arbitrary polynomial degree to remain general and extensible.
In general, model parameters such as Wilson coefficients can be complex. In such cases, observables may depend separately on the real and imaginary parts of the parameters. For example, a quadratic monomial $C_i C_j$ must be interpreted as a combination of terms such as $\mathrm{Re}(C_i)\mathrm{Re}(C_j)$, $\mathrm{Re}(C_i)\mathrm{Im}(C_j)$, etc. The format explicitly supports such terms via a real-imaginary decomposition.

Although we use the term `observable' to refer to a quantity whose prediction is a single real number, in many cases the objects of interest are vector-valued quantities, i.e.\ sets of observables in the aforementioned sense. Examples include binned distributions and angular observables. In such cases, each bin or component is treated as a separate observable, and the predictions of the entire set of observables are encoded via vector-valued polynomial coefficients. The data format accommodates this naturally via array structures compatible with numerical data structures such as \texttt{numpy} or \texttt{Mathematica} arrays.

Some observables are not themselves polynomials in the parameters but are functions of one or more polynomial expressions. Examples include ratios of decay widths or angular observables formed from linear combinations of angular coefficients. The data format supports such derived observables by allowing them to be defined in terms of functional combinations of underlying polynomials.

Finally, the treatment of uncertainties and correlations in this format assumes that the observable is expressed as a polynomial in the model parameters. In many cases, the observable itself is already a polynomial -- for example, cross sections or partial decay widths derived from squared amplitudes.
In the case where the observable is defined as a function of one or more polynomials, it is expanded in the model parameters to yield a single polynomial expression, typically truncated at second order.
This expansion defines the level at which parameter-dependent uncertainties and correlations are interpreted, and provides a consistent and efficient framework for uncertainty propagation~\cite{Altmannshofer:2021qrr}.

The proposed data format is based on the \texttt{JSON} standard, a lightweight, text-based data format that is both human-readable and natively supported in most programming languages and computer algebra systems, particularly in \texttt{python} and \texttt{Mathematica}, which are used extensively in the particle physics community. It provides a simple and transparent way to encode central values, uncertainties and correlations as well as all relevant metadata in a structured, hierarchical format. Specific design goals include:
\begin{itemize}
    \item general applicability to any observable that can be expressed in terms of a polynomial in model parameters, including but not limited to EFT Wilson coefficients;
    \item support for arbitrary polynomial degree, enabling the inclusion of higher-order terms such as those arising from dimension-eight EFT operators;
    \item support for sectors of related observables, such as binned distributions, angular observables, correlated decay channels or, more generally, observables that depend on the same set of parameters;
    \item support for observables defined as functions of one or multiple polynomial parameterisations, such as normalised distributions or ratios of decay widths;
    \item explicit encoding of all assumptions, such as parameter basis definitions, input parameter values, renormalisation scale, and operator normalisations, to ensure reproducibility;
    \item for interfacing to existing tools in EFT applications, support for the commonly-used Wilson Coefficient exchange format (WCxf)~\cite{Aebischer:2017ugx}, which encodes generic EFT parametrisations, defining the operator basis and normalisations in an unambiguous way;
    \item consistent treatment of theoretical uncertainties from various sources, including support for parameter-dependent uncertainties;
    \item support for correlated uncertainties between observables, including the possibility of parameter-dependent correlations.
\end{itemize}
In the following sections, we will present the proposed structure of this data format in detail and illustrate its use with several concrete examples. Our aim is to provide a general framework that can be readily adopted by both theorists and experimentalists for the publication, exchange, and application of theoretical predictions, with a special emphasis on EFT-based parametrisations.
\Cref{sec:formalism} outlines the general formalism for data files describing predictions that can be expressed as a function of polynomials in the model parameters, as well as the associated uncertainties. \Cref{sec:overview} presents an overview of the \texttt{JSON} structure used to encode the polynomial data, and introduces the two intended modes of use for the data format: \emph{single-polynomial} mode and \emph{function-of-polynomials} mode. \Cref{specification-of-fields-in-the-popxf-json-format} provides a detailed specification of the fields in the \texttt{POPxf} format.
\Cref{sec:correlation_files} describes the structure of the separate \texttt{POPxf} correlation files that are used to store correlation coefficients of correlated observables.
Finally,~\cref{sec:conclusions} briefly summarises and concludes.

The concrete specification of the \texttt{POPxf} data format is hosted in a public repository at \url{https://github.com/pop-xf} in terms of \texttt{JSON} schemas, which enable the automated validation of \texttt{POPxf} files by existing \texttt{JSON} schema validators (allowed/required fields, data types, etc.).
The repository also includes a collection of example files and several useful tools. Among these is an implementation of a lightweight validator that performs additional consistency checks on, \emph{e.g.}, lengths of various fields depending on the specified mode of use.

For impatient readers who are keen to get started immediately, we recommend taking a brief look at~\cref{sec:overview}, \cref{fig:format}, and the explicit examples given in~\cref{app:examples}.
\section{General Formalism}\label{sec:formalism}

\subsection{Data Files}

We consider polynomial observable predictions defined within a set of $N$ data files, which are labelled by an index $n\in[1,N]$.
The terms \emph{observable} and \emph{polynomial} refer to real-valued scalar quantities.
The data file with index $n$ defines:
\begin{itemize}
 \item A basis of $S^{(n)}$ \emph{parameters} $C_s^{(n)}$ on which the predictions depend (in the case of EFT predictions, the parameters are the Wilson coefficients), and which are labelled by indices $s\in[1,S^{(n)}]$. Each parameter can be real or complex. Denoting the number of real and complex parameters by $S_{\mathbb{R}}^{(n)}$ and  $S_{\mathbb{C}}^{(n)}$, respectively, with $S^{(n)}=S_{\mathbb{R}}^{(n)}+S_{\mathbb{C}}^{(n)}$, we denote the real-valued, real and imaginary parts of the parameters as $\hat C_r^{(n)}$, which are labelled by indices $r\in[1,R^{(n)}]$ with $R^{(n)}=S_{\mathbb{R}}^{(n)}+2\, S_{\mathbb{C}}^{(n)}$. We group the $\hat C_r^{(n)}$ into the real-valued vector $\vec C^{(n)}$. Note that this vector is in general \emph{not} a vector of the (potentially complex) parameters $C_s^{(n)}$, but a vector of their real and imaginary parts $\hat C_r^{(n)}$, i.e.~the components of the vector are given by $[\vec C^{(n)}]_r = \hat C_r^{(n)}$. For instance, for a set of two complex parameters $C_1,\,C_2$, the vector would be $[\mathrm{Re}(C_1),\mathrm{Im}(C_1), \mathrm{Re}(C_2), \mathrm{Im}(C_2)]$.

 \item A set of $K^{(n)}$ \emph{polynomials} $P_{k}^{(n)}$, which are labelled by indices $k\in[1,K^{(n)}]$. The polynomials are specified in terms of \emph{polynomial coefficients} $\vec p_{k}^{\,(n)}$ (see below).
 \item A set of $M^{(n)}$ \emph{observable predictions} $O_{m}^{(n)}$, which are labelled by indices $m\in[1,M^{(n)}]$. The observable predictions are defined in terms of $M^{(n)}$ functions of the polynomials~$P_{k}^{(n)}$, which we denote as \emph{observable expressions} $E_{m}^{(n)}$,
 \begin{equation}
   O_{m}^{(n)} = E_{m}^{(n)}\left(P_1^{(n)}, P_2^{(n)}, ..., P_{K^{(n)}}^{(n)}\right)\,.
 \end{equation}
 In many practical examples, all observables defined in a given data file are themselves polynomials, such that the number of polynomials and observables coincide, $K^{(n)}=M^{(n)}$, and each observable expression $E_{m}^{(n)}$ is a trivial identity function of a single polynomial labelled by $k=m$,
 \begin{equation}
  O_{m}^{(n)}  = E_{m}^{(n)}\left(P_{m}^{(n)}\right) = P_{m}^{(n)}\,.
 \end{equation}
 In this special case, no observable expressions have to be specified, and the observable predictions are directly given in terms of the polynomial coefficients $\vec p_{m}^{\,(n)}$, which are then denoted as \emph{observable coefficients} $\vec o_{m}^{\,(n)}=\vec p_{m}^{\,(n)}$.
 But in the general case, observables are given by non-trivial functions of multiple polynomials.
\end{itemize}%

\subsection{Polynomials}\label{sec:formalism_poly_pred}
Each polynomial $P_{k}^{(n)}$ can be expressed as a scalar product of two vectors: a vector of \emph{polynomial coefficients} $\vec p_{k}^{\ (n)}$, and a vector of \emph{parameter monomials} $\vec V^{(n)}$,
\begin{equation}
 P_{k}^{(n)} = \vec p_{k}^{\ (n)} \cdot \vec V^{(n)}\,.
\end{equation}
For polynomials of degree~$d$, the vector $\vec V^{(n)}$ is given by the set of all monomials up to degree~$d$ formed from the components of $\vec C^{(n)}$.

In the following, we will focus on second-order polynomials\footnote{In principle, the format can support higher degree polynomials, with a corresponding generalisation of the equations that follow. For example the vectors $\vec{V}^{(n)}$ can be extended by higher degree monomials of the $\vec{C}^{(n)}$ parameters, with a corresponding extension of the $\vec{p}^{\ (n)}$ coefficients. }, for which  $\vec V^{(n)}$ takes the form
\begin{equation}
 \vec V^{(n)} = \begin{pmatrix}
 1 \\
 \vec C^{(n)} \\
 {\rm vech}( \vec C^{(n)} \otimes \vec C^{(n)})
 \end{pmatrix}\,,
\end{equation}
where $\rm vech$ denotes half-vectorization,\footnote{%
The half-vectorization of a symmetric $n\times n$ matrix $A$ is defined as the column vector with $n(n+1)/2$ components obtained by stacking the lower triangular columns of $A$ on top of each other:
\[
 {\rm vech}(A) = (A_{1 1}, A_{2 1}, \ldots, A_{n 1}, A_{2 2}, A_{3 2}, \ldots, A_{n 2}, \ldots, A_{nn})^T
\]
Applying $\rm vech$ to the outer product of a vector $\vec v$ with itself, $\vec v \otimes \vec v$, yields a vector of all unique degree-two monomials formed from the components of $\vec v$,
\[
{\rm vech}( \vec v \otimes \vec v) = (v_{1}^2, v_{1}v_{2}, v_{1}v_{3}, ..., v_{1}v_{n}, v_{2}^2, v_{2}v_{3}, ..., v_{2}v_{n}, ...,v_{n-1}^2, v_{n-1}v_n, v_{n}^2)^T\,.
\]
}
and we have split $\vec V^{(n)}$ into three parts corresponding to the parameter monomials of degrees zero, one, and two.
We denote the components of the vector $\vec V^{(n)}$ as $[\vec V^{(n)}]_\alpha$ labelled by indices $\alpha\in [0,A^{(n)}]$, where in the case of second-degree polynomials $A^{(n)}$ is given by
\begin{equation}\label{eq:num_parameter_monomials}
 A^{(n)} = R^{(n)} + R^{(n)}(R^{(n)}+1)/2 = R^{(n)}(R^{(n)}+3)/2\,,
\end{equation}
such that
\begin{equation}
 [\vec V^{(n)}]_\alpha =
 \begin{cases}
  1 & {\rm if\ }  \alpha = 0 \\
  [\vec C^{(n)}]_\alpha & {\rm if\ }  \alpha \in [1,R^{(n)}] \\
  [{\rm vech}( \vec C^{(n)} \otimes \vec C^{(n)})]_{\alpha-R^{(n)}} & {\rm if\ }  \alpha \in [R^{(n)}+1, A^{(n)}] \\
 \end{cases}\,.
\end{equation}
A vector of polynomial coefficients can then be written as
\begin{equation}\def\arraystretch{1.5}\label{eq:poly_coeffs_vector}
\vec p_{k}^{\ (n)} = \begin{pmatrix}
 a_{k}^{(n)} \\
 \vec b_{k}^{\,(n)} \\
 \vec c_{k}^{\ (n)}
 \end{pmatrix}\,,
\end{equation}
such that the polynomials can be given by
\begin{equation}\label{eq:poly_pred}
 P_{k}^{(n)}
 =
 \vec p_{k}^{\ (n)} \cdot \vec V^{(n)}
 =
 a_{k}^{(n)}
 +
 \vec b_{k}^{\,(n)} \cdot \vec C^{(n)}
 +
 \vec c_{k}^{\ (n)} \cdot {\rm vech}( \vec C^{(n)} \otimes \vec C^{(n)})\,,
\end{equation}
i.e.\ $a_{k}^{(n)}$, $\vec b_{k}^{\,(n)}$, and $\vec c_{k}^{\ (n)}$ contain the polynomial coefficients that multiply parameter monomials of degrees zero, one, and two, respectively.
In other words, $a_{k}^{(n)}$ is the parameter-independent constant term, while $\vec b_{k}^{\,(n)}$ and $\vec c_{k}^{\ (n)}$ contain the coefficients of the terms linear and quadratic in the parameters.

The numerical values of the polynomial coefficients $\vec p_{k}^{\ (n)}$ uniquely define the polynomials $P_{k}^{(n)}$ as functions of the parameters $\vec C^{(n)}$.

\subsection{Observable Predictions}\label{sec:formalism_obs_pred}

We distinguish between two modes of use for specifying observable predictions:
\begin{enumerate}
 \item The \emph{function-of-polynomials} (FOP) mode defines observable predictions $O_{m}^{(n)}$ in the general case, in which they are given by non-trivial observable expressions $E_{m}^{(n)}$, which are arbitrary functions of polynomials $P_{k}^{(n)}$ (note that both the $O_{m}^{(n)}$ and the $P_{k}^{(n)}$ are real-valued scalar quantities),
 \begin{equation}
   O_{m}^{(n)}\big|_{\rm FOP} = E_{m}^{(n)}\left(P_1^{(n)}, P_2^{(n)}, ..., P_{K^{(n)}}^{(n)}\right)\,.
 \end{equation}

 \item The \emph{single-polynomial} (SP) mode can be used if all observables in a given data file are themselves polynomials, i.e.\ if their observable expressions $E_{m}^{(n)}$ are all trivial identity functions. In this case, no observable expressions have to be specified, and the observable predictions have the same structure as the polynomials defined above. In SP mode, we express each observable prediction $O_{m}^{(n)}$ directly as a scalar product of \emph{observable coefficients} $\vec o_{m}^{\ (n)}$, and the vector of parameter monomials $\vec V^{(n)}$ defined in~\cref{sec:formalism_poly_pred},
\begin{equation}
 O_{m}^{(n)}\big|_{\rm SP}  = \vec o_{m}^{\ (n)} \cdot \vec V^{(n)}\,.
\end{equation}
Focusing again on second-order polynomials, the observable coefficients can be written as
\begin{equation}\def\arraystretch{1.5}
\vec o_{m}^{\ (n)} = \begin{pmatrix}
 a_{m}^{(n)} \\
 \vec b_{m}^{\,(n)} \\
 \vec c_{m}^{\ (n)}
 \end{pmatrix}\,,
\end{equation}
such that the observable predictions are given by
\begin{equation}
 O_{m}^{(n)}\big|_{\rm SP}
 =
 \vec o_{m}^{\ (n)} \cdot \vec V^{(n)}
 =
 a_{m}^{(n)}
 +
 \vec b_{m}^{\,(n)} \cdot \vec C^{(n)}
 +
 \vec c_{m}^{\ (n)} \cdot {\rm vech}( \vec C^{(n)} \otimes \vec C^{(n)})\,,
\end{equation}
analogous to the polynomials in Eq.~\eqref{eq:poly_pred}.

\end{enumerate}
While in SP mode an observable prediction is exactly given by a single polynomial, in FOP mode it can be approximated by a single polynomial if all parameters are small, $\vec C^{(n)}\ll 1$ and  Taylor expansion around $\vec C^{(n)} = \vec 0$ is possible.
Such an approximation will be needed for our treatment of parameter-dependent uncertainties discussed in Section~\ref{sec:uncertainties}, which relies on each observable being expressed as a single polynomial.
If all parameters are small,
a consistent truncation of the multivariate expansion can be obtained by defining the small quantity $\epsilon\ll1$ as the magnitude of the largest parameter, $\epsilon=\max(|[\vec C^{n}]_r|)$, and expressing
\begin{equation}
 \vec C^{(n)} = \epsilon\,\vec{\mathcal C}^{\,(n)}
 \qquad
 \text{with}
 \qquad
 \epsilon\ll 1
 \qquad
 \text{and}
 \qquad
 |[\vec{\mathcal C}^{n}]_r| \leq 1\,.
\end{equation}
For second-order polynomials, the vector of parameter monomials, $\vec V^{(n)}$, thus contains terms up to $\mathcal{O}(\epsilon^2)$, and the observable predictions in the general case can be expanded as
\begin{equation}\label{eq:FOP_expansion}
 O_{m}^{(n)}\big|_{\rm FOP} = \vec o_{m}^{\ (n)} \cdot \vec V^{(n)} + \mathcal{O}(\epsilon^3)
 =
 a_{m}^{\prime(n)}
 + \epsilon
 \left(
 \vec b_{m}^{\,\prime(n)} \cdot \vec{\mathcal C}^{\,(n)}
 \right)
 +
 \epsilon^2
 \left(
 \vec c_{m}^{\ \prime(n)} \cdot {\rm vech}( \vec{\mathcal C}^{\,(n)} \otimes \vec{\mathcal C}^{\,(n)})
 \right)+ \mathcal{O}(\epsilon^3)
 \,.
\end{equation}
This approximated observable prediction has exactly the same form as the SP mode prediction, and we have defined the corresponding observable coefficients $\vec o_{m}^{\ (n)}$ in terms of the primed quantities $a_m^{\prime(n)}$, $\vec b_m^{\,\prime(n)}$, and $\vec c_m^{\ \prime(n)}$.
They can be obtained from the observable expressions $E^{(n)}_m$ and the polynomial coefficients $a_k^{(n)}$, $\vec b_k^{\,(n)}$, and $\vec c_k^{\ (n)}$ through~\cite{Altmannshofer:2021qrr}
\begin{equation}\label{eq:FOP_expansion_coefficients}
\begin{aligned}
 a_{m}^{\prime(n)} &= [G_{m}^{(n)}]\,,
 \\
 \vec b_{m}^{\,\prime(n)} &= \sum_{k_1} [G_{m}^{(n)}]_{k_1}\,\vec b_{k_1}^{\,(n)}\,,
 \\
 \vec c_{m}^{\ \prime(n)} &= \sum_{k_1} [G_{m}^{(n)}]_{k_1}\,\vec c_{k_1}^{\ (n)} +\frac{1}{2} \sum_{k_1,k_2} [G_{m}^{(n)}]_{k_1 k_2}\,D_{R^{(n)}}^T\,{\rm vec}( \vec b_{k_1}^{\,(n)} \otimes \vec b_{k_2}^{\,(n)})\,,
\end{aligned}
\end{equation}
where $D_{R^{(n)}}^T$ is the transpose of the duplication matrix\footnote{
The duplication matrix $D_{n}$ is the unique $n^2 \times n(n+1)/2$ matrix that for any symmetric $n\times n$ matrix A relates its vectorization ${\rm vec}(A)$ and its half-vectorization ${\rm vech}(A)$ by~\cite{doi:10.1137/0601049}
\begin{equation}
 {\rm vec}(A) = D_n\, {\rm vech}(A)\,.
\end{equation}
}
with $R^{(n)}$ the number of parameters,
and the coefficients $[G_{m}^{(n)}]_{k_1 k_2\ldots k_\ell}$ of the multivariate Taylor expansion are defined as the $\ell$-th derivatives of the observable expressions $E_{m}^{(n)}$ evaluated at $\epsilon=0$, i.e.\ for $P_k^{(n)}= a_k^{(n)}$:
\begin{equation}
[G_{m}^{(n)}]_{k_1 k_2\ldots k_\ell} = \frac{\partial^\ell E_{m}^{(n)}\left(P_1^{(n)}, P_2^{(n)}, ..., P_{K^{(n)}}^{(n)}\right)}{\partial_{P_{k_1}^{(n)}}\,\partial_{P_{k_2}^{(n)}}\ldots \partial_{P_{\ell}^{(n)}} }\Bigg|_{
P_1^{(n)}=a_1^{(n)},\ P_2^{(n)}=a_2^{(n)},...,\ P_{K^{(n)}}^{(n)}=a_{K^{(n)}}^{(n)}
}\,,
 \end{equation}
 such that the $\ell=0$ coefficient is
\begin{equation}
 [G_{m}^{(n)}] = E_{m}^{(n)}\left(a_1^{(n)}, a_2^{(n)}, ..., a_{K^{(n)}}^{(n)}\right)\,.
\end{equation}
In the case where the $E_{m}^{(n)}$ are identity functions, $E_{m}^{(n)}\big(P_m^{(n)}\big)=P_m^{(n)}$, we recover the observable prediction in SP mode, with $[G_{m}^{(n)}] = a_m^{(n)}$, $[G_{m}^{(n)}]_{k_1}=\delta_{m,k_1}$, and $[G_{m}^{(n)}]_{k_1\ldots k_\ell}=0$ for $\ell\geq 2$.
A slightly less trivial example is a ratio of two polynomials, $E_{m}^{(n)}\big(P_1^{(n)},P_2^{(n)}\big)=P_1^{(n)}/P_2^{(n)}$. In this case one obtains
\begin{equation}
    \begin{aligned}
    [G_{m}^{(n)}] &= a_1^{(n)}/a_2^{(n)}\,,\quad
    &[G_{m}^{(n)}]_1 &= 1/a_2^{(n)}\,,\quad
    &[G_{m}^{(n)}]_2 &= -a_1^{(n)}/(a_2^{(n)})^2\,,\\
    [G_{m}^{(n)}]_{11} &= 0\,,\quad
    &[G_{m}^{(n)}]_{12} = [G_{m}^{(n)}]_{21} &= -1/(a_2^{(n)})^2\,,\quad
    &[G_{m}^{(n)}]_{22} &= 2\,a_1^{(n)}/(a_2^{(n)})^3\,.
    \end{aligned}
\end{equation}

\subsection{Observable Uncertainties and Correlations\label{sec:uncertainties}}

We consider Gaussian uncertainties in observable predictions that can be both correlated and depend on the $\vec C^{(n)}$ parameters.\footnote{The assumption of Gaussianity is required in order for the distribution of the numerical predictions to be unambiguously characterized by the uncertainty value. Alternative distributions, obtained e.g. via replica simulations, could be reproduced by employing a separate file for each replica.}
In SP mode -- when the observables are themselves polynomials -- the uncertainties and correlations of the observable predictions $O_m^{(n)}|_{\rm SP}$ can be expressed analytically in terms of the uncertainties and correlations of their observable coefficients $\vec o_m^{\ (n)}$ (see below). In FOP mode -- when the observables are given by arbitrary functions of polynomials -- for the purposes of uncertainties we only consider the case of small parameters, which allows us to expand the observable predictions $O_m^{(n)}|_{\rm FOP}$ as in Eq.~\eqref{eq:FOP_expansion} with approximate observable coefficients $\vec o_m^{\ (n)}$ given by Eq.~\eqref{eq:FOP_expansion_coefficients}. This allows us to treat uncertainties and correlations in the same way for both SP and FOP observables, defined through the uncertainties and correlations of the exact (in SP mode) or approximate (in FOP mode) observable coefficients $\vec o_m^{\ (n)}$.

In the following, we express the uncertainties of the observable coefficients $\vec o_m^{\ (n)}$ by the vector
\begin{equation}
\vec {\sigma}_m^{\ (n)}
\end{equation}
and the correlations between observable coefficients $\vec o_m^{\ (n)}$ (of observable $m$ in data file $n$) and $\vec o_{m^\prime}^{\ (n^\prime)}$ (of observable $m^\prime$ in data file $n^\prime$) by the correlation matrix
\begin{equation}
 \rho_{m m^\prime}^{(n n^\prime)}\,.
\end{equation}
In addition, we denote the components of the vector $\vec {\sigma}_m^{\ (n)}$ as
\begin{equation}
 [\vec {\sigma}_m^{\ (n)}]_\alpha
\end{equation}
and the components of the matrix $\rho_{m m^\prime}^{(n n^\prime)}$ as
\begin{equation}
 [\rho_{m m^\prime}^{(n n^\prime)}]_{\alpha\alpha^\prime}
\end{equation}
with $\alpha \in [0,A^{(n)}]$ and $\alpha^\prime \in [0,A^{(n^\prime)}]$, where, as in Eq.~\eqref{eq:num_parameter_monomials}, for second-order polynomials $A^{(n)} = R^{(n)}(R^{(n)}+3)/2$ and $A^{n^\prime}=R^{(n^\prime)}(R^{(n^\prime)}+3)/2$.

The correlation matrix $\rho_{m m^\prime}^{(n n^\prime)}$ and the uncertainties $\vec {\sigma}_m^{\ (n)}$ and $\vec {\sigma}_{m^\prime}^{\ (n^\prime)}$ can be combined into a covariance matrix for the observable coefficients $\vec o_m^{\ (n)}$ and $\vec o_{m^\prime}^{\ (n^\prime)}$, which we denote as
\begin{equation}
\tilde \Sigma_{m m^\prime}^{(n n^\prime)}\,.
\end{equation}
The components of this matrix are given by
\begin{equation}\label{eq:Sigma_tilde_components}
 [\tilde \Sigma_{m m^\prime}^{(n n^\prime)}]_{\alpha \alpha^\prime} =
 [\vec{\sigma}_m^{\ (n)}]_\alpha\,
 [\rho_{m m^\prime}^{(n n^\prime)}]_{\alpha\alpha^\prime}\,
 [\vec{\sigma}_{m^\prime}^{\ (n^\prime)}]_{\alpha^\prime}\,,
\end{equation}
i.e.\ the covariance matrix for the observable coefficients can be expressed in terms of the observable coefficients' uncertainties and correlations.

We allow different sources of uncertainties and correlations to be defined separately (e.g.\ from Monte Carlo statistics, remaining scale-dependence, parton distribution functions, etc.). In this case, separate uncertainties
\begin{equation}
 [\vec{\sigma}_m^{\ (n)}]_\alpha^{(\text{source})}
\end{equation}
and correlations
\begin{equation}
 [\rho_{m m^\prime}^{(n n^\prime)}]_{\alpha\alpha^\prime}^{(\text{source})}
\end{equation}
are defined for each source. The components of the total covariance matrix for the observable coefficients are then obtained by summing the individual covariance matrices for each source,
\begin{equation}\label{eq:Sigma_tilde_components_sources}
 [\tilde \Sigma_{m m^\prime}^{(n n^\prime)}]_{\alpha \alpha^\prime} =
 \sum_{(\text{source})}
 [\tilde \Sigma_{m m^\prime}^{(n n^\prime)}]_{\alpha \alpha^\prime}^{(\text{source})}
 =
 \sum_{(\text{source})}
 [\vec{\sigma}_m^{\ (n)}]_\alpha^{(\text{source})}\,
 [\rho_{m m^\prime}^{(n n^\prime)}]_{\alpha\alpha^\prime}^{(\text{source})}\,
 [\vec{\sigma}_{m^\prime}^{\ (n^\prime)}]_{\alpha^\prime}^{(\text{source})}\,.
\end{equation}

The uncertainties and correlations of the observable predictions $O_m^{(n)}$ and $O_{m^\prime}^{(n^\prime)}$ can be expressed in terms of a covariance matrix of the observables, $\Sigma$, which can be written as a block matrix, with the block $\Sigma^{(n n^\prime)}$ corresponding to the covariances between data files $n$ and $n^\prime$,
\begin{equation}
 \Sigma =
 \begin{pmatrix}
\Sigma^{(1 1)} & \Sigma^{(1 2)} & ... & \Sigma^{(1 N)}\\
\Sigma^{(2 1)} & \Sigma^{(2 2)} & ... & \Sigma^{(2 N)}\\
\vdots & \vdots & \ddots & \vdots \\
\Sigma^{(N 1)} & \Sigma^{(N 2)} & ... & \Sigma^{(N N)}\\
\end{pmatrix}\,.
\end{equation}
The components of the block $\Sigma^{(n n^\prime)}$ can be labelled by indices $m$ and $m^\prime$ and correspond to the covariances of observable predictions $O_m^{(n)}$ and $O_{m^\prime}^{(n^\prime)}$ and we denote them as
\begin{equation}
 [\Sigma^{(n n^\prime)}]_{m m^\prime}\,.
\end{equation}
In case of a single data file, $N=1$, the components of the full covariance matrix of the observable predictions $O_m^{(1)}$ and $O_{m^\prime}^{(1)}$ are simply
\begin{equation}
 \Sigma_{m m^\prime}
 =
 [\Sigma^{(1 1)}]_{m m^\prime}\,.
\end{equation}

\subsubsection{Parameter-independent Uncertainties and Correlations}

When considering uncertainties and correlations of observable predictions, it is common practice to make the assumption that their parameter dependence can be neglected, and that they are well approximated by the uncertainties and correlations of the parameter-independent constant term.
This is in particular the case in EFT applications where the parameters are new physics Wilson coefficients. In this case, the constant term is nothing but the SM prediction, and considering only the SM uncertainties and correlations is often an excellent approximation.

Neglecting the parameter-dependence, the covariances of observable predictions $O_m^{(n)}$ and $O_{m^\prime}^{(n^\prime)}$ are given by
\begin{equation}
 [\Sigma^{(n n^\prime)}]_{m m^\prime} =
 [\tilde \Sigma_{m m^\prime}^{(n n^\prime)}]_{00}
\end{equation}
where $[\tilde \Sigma_{m m^\prime}^{(n n^\prime)}]_{00}$ are the covariances of the constant terms in the observable predictions $O_m^{(n)}$ and $O_{m^\prime}^{(n^\prime)}$  (the SM covariances in new physics EFT applications),
defined by \cref{eq:Sigma_tilde_components} in case of a single source of uncertainties and \cref{eq:Sigma_tilde_components_sources} in case of multiple sources.

\subsubsection{Parameter-dependent Uncertainties and Correlations}

If the parameter-dependence of the uncertainties and correlations cannot be neglected, it is still possible to express the covariances $[\Sigma^{(n n^\prime)}]_{m m^\prime}$ of observable predictions $O_m^{(n)}$ and $O_{m^\prime}^{(n^\prime)}$ in terms of the observable coefficients' covariances $[\tilde \Sigma_{m m^\prime}^{(n n^\prime)}]_{\alpha {\alpha^\prime}}$ and the parameter monomials $[\vec V^{(n)}]_\alpha$ and $[\vec V^{(n^\prime)}]_{\alpha^\prime}$, as discussed in~\cite{Altmannshofer:2021qrr},
\begin{equation}
 [\Sigma^{(n n^\prime)}]_{m m^\prime} =
 \sum_{\alpha,{\alpha^\prime}}
 [\vec V^{(n)}]_\alpha\,
 [\tilde \Sigma_{m m^\prime}^{(n n^\prime)}]_{\alpha {\alpha^\prime}}\,
 [\vec V^{(n^\prime)}]_{\alpha^\prime}
 \,,
\end{equation}
where $[\tilde \Sigma_{m m^\prime}^{(n n^\prime)}]_{\alpha {\alpha^\prime}}$ is defined by \cref{eq:Sigma_tilde_components} in case of a single source of uncertainties and \cref{eq:Sigma_tilde_components_sources} in case of multiple sources.

\section{Overview of the \texttt{POPxf} \texttt{JSON} Format}\label{sec:overview}

\begin{figure}[!ht]
    \centering
    \includegraphics[trim={2.3cm 1.1cm 2.2cm 1.1cm},clip,width=0.8\textwidth]{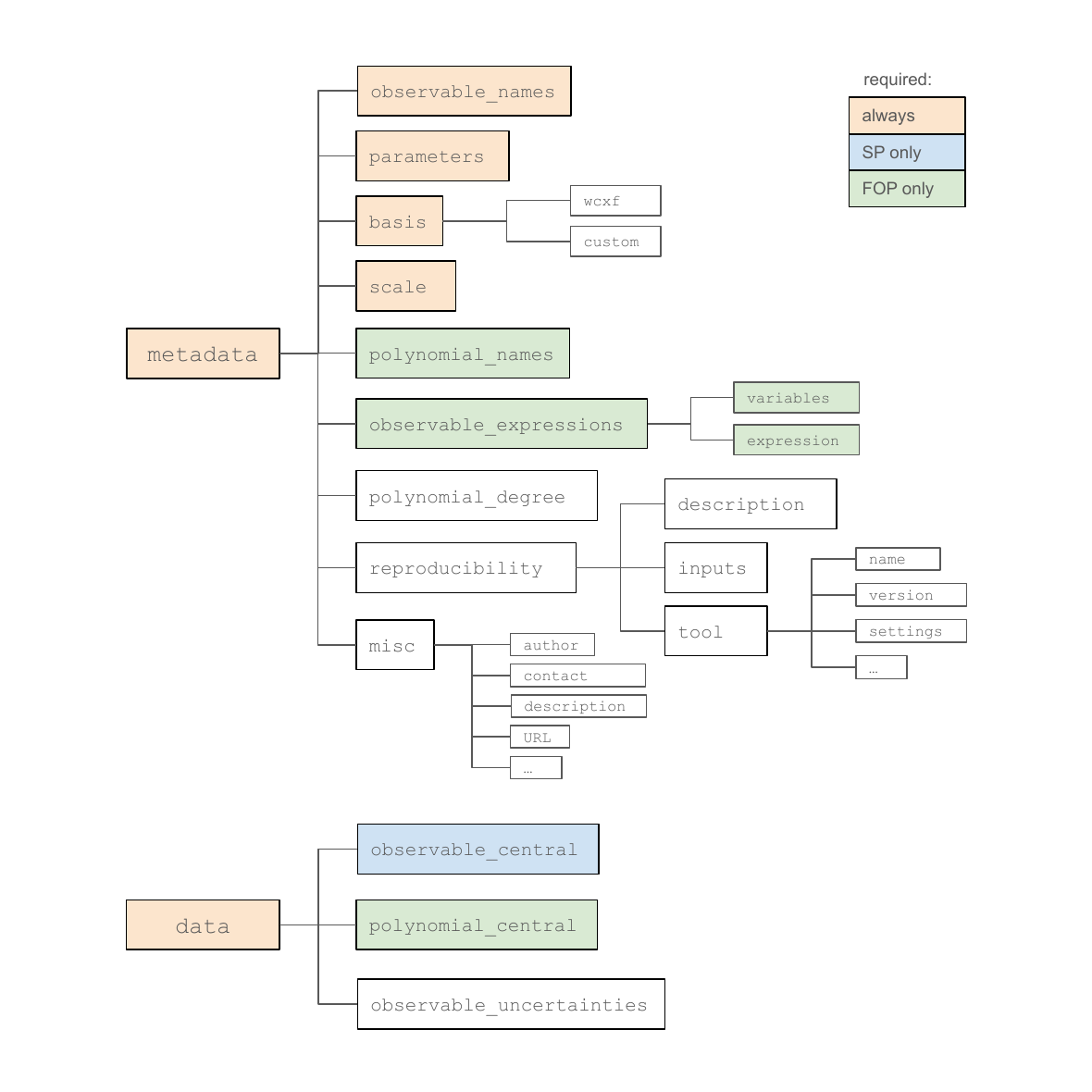}
    \caption{Graphical representation of the \texttt{JSON} structure of the \texttt{POPxf} data format for the two top-level fields, \texttt{metadata} and \texttt{data}. Coloured boxes indicate required fields (blue and green for SP and FOP modes, respectively, orange for both). White boxes indicate fields that are always optional.}
    \label{fig:format}
\end{figure}

This section provides a technical overview of the \texttt{JSON} structure used to encode theoretical predictions of observables that can be expressed as polynomials in model parameters. It introduces the overall layout and key structural concepts, including the separation between contextual metadata and numerical prediction data, and the distinction between single-polynomial and function-of-polynomials use cases. \Cref{fig:format} shows a graphical representation of the fields in the \texttt{POPxf} data structure.

\subsection{Top-Level Structure}
Predictions of a set of observables are encoded as a single \texttt{JSON} object containing exactly three required top-level fields:
\begin{itemize}
    \item \textbf{\texttt{\$schema}}: Keyword that declares the \texttt{JSON} schema specification used by the file. This allows identifying a given \texttt{JSON} file as a \texttt{POPxf} file and validating its structure. The value must be \texttt{"https://json.schemastore.org/popxf-1.0.json"} for the first version of this format and the version number will be incremented if the format is extended in the future;
    \item \textbf{\texttt{metadata}}: contextual and structural information required to interpret and reproduce the prediction;
    \item \textbf{\texttt{data}}: numerical information representing observables in terms of their polynomial coefficients and uncertainties on these.
\end{itemize}
No additional top-level keys are defined or permitted in the current schema.

\subsection{Two Modes of Use}\label{sec:twomodes}
As introduced in~\cref{sec:formalism_obs_pred}, the format distinguishes between two structurally distinct ways to define an observable prediction.

\subsubsection*{1. Function-of-polynomials mode.}
The observables are defined as functions of one or more named polynomials. This most general case allows for non-polynomial structures such as ratios, products, or square roots of polynomial components. In this case, the following fields are used:
\begin{itemize}
    \item \textbf{\texttt{metadata.polynomial\_names} (\emph{required})}: array of polynomial names;
    \item \textbf{\texttt{metadata.observable\_expressions} (\emph{required})}: Python-like expressions defining how each observable is computed from the named polynomials;
    \item \textbf{\texttt{data.polynomial\_central} (\emph{required})}: central values of polynomial coefficients, $\vec p_{k}$, for each named polynomial;
    \item \textbf{\texttt{data.observable\_uncertainties} (\emph{optional})}: uncertainties, $\vec {\sigma}_m$, on the polynomial coefficients of each observable after expanding it in the model parameters (see~\cref{sec:uncertainties});
    \item \textbf{\texttt{data.observable\_central} (\emph{optional})}: central values, $\vec o_{m}$, of polynomial coefficients for each observable after expanding it in the model parameters.
    In FOP mode, this field is optional and is not intended to be used to compute the actual central values of the observables. It may be included to allow numerical comparison between the full observable expression and its expansion in the model parameters, or to enable switching between function-of-polynomials mode and single-polynomial mode.
\end{itemize}
The presence of any of the three required fields of function-of-polynomials mode will assume the use of this mode such that the other two are also required.

\subsubsection*{2. Single-polynomial mode.}
Every observable is directly defined by a single polynomial in the model parameters. In this special but commonly-used case, the following fields are used:
\begin{itemize}
    \item \textbf{\texttt{data.observable\_central} (\emph{required})}: central values of the observable coefficients, $\vec o_{m}=\vec p_{m}$, the polynomial coefficients of the observable;
    \item \textbf{\texttt{data.observable\_uncertainties} (\emph{optional})}: uncertainties on these observable coefficients, $\vec {\sigma}_m$;
\end{itemize}
The \texttt{metadata} fields \texttt{polynomial\_names} and \texttt{observable\_expressions} are not used in single-polynomial mode.

As discussed in~\cref{sec:uncertainties}, in both modes, uncertainties and correlations are interpreted based on a polynomial expansion of each observable in the model parameters, truncated at the order defined by the \texttt{metadata} field \texttt{polynomial\_degree} (see next section).

\subsection{Structure of \texttt{metadata}}

The \texttt{metadata} field contains all contextual and structural information required to interpret the predictions. Its subfields are listed below. A full specification of each subfield, including its structure, format, input types and allowed values, is provided in~\cref{specification-of-fields-in-the-popxf-json-format}.

\begin{itemize}
    \item \textbf{\texttt{observable\_names} (\emph{required})}: array of observable names;
    \item \textbf{\texttt{parameters} (\emph{required})}: array of model parameters (e.g.\ names of Wilson coefficients);
    \item \textbf{\texttt{basis} (\emph{required})}: definition of the parameter basis, e.g.\ the operator basis in an EFT;
    \item \textbf{\texttt{scale} (\emph{required})}: renormalisation scale at which the parameters are defined;
    \item \textbf{\texttt{polynomial\_names} (\emph{required} in FOP mode)}: array of polynomial names;
    \item \textbf{\texttt{observable\_expressions} (\emph{required} in FOP mode)}: Python-like expressions defining how each observable is computed from the named polynomials;
    \item \textbf{\texttt{polynomial\_degree} (\emph{optional})}: the order of truncation in the parameter expansion (default: 2);
    \item \textbf{\texttt{reproducibility} (\emph{optional})}: information needed to reproduce the predictions. Array in which each element may include the following fields:
    \begin{itemize}
    \item \textbf{\texttt{inputs} (\emph{optional})}: numerical values of input parameters that have been used to compute the polynomial coefficients, with optional uncertainties and correlations;
    \item \textbf{\texttt{tool} (\emph{optional})}: information about the tools/methods used to generate the predictions;
    \item \textbf{\texttt{description} (\emph{optional})}: text field with a summary of how the predictions were obtained.
    \end{itemize}
    \item \textbf{\texttt{misc} (\emph{optional})}: free-form documentation or provenance metadata.
\end{itemize}

\subsection{Structure of \texttt{data}}

The \texttt{data} field contains the numerical predictions. Its subfields, already described in~\cref{sec:twomodes}, are listed below. A full specification of each subfield, including its structure, format, input types and allowed values, is provided in~\cref{specification-of-fields-in-the-popxf-json-format}.

\begin{itemize}
    \item \textbf{\texttt{polynomial\_central} (\emph{required} in FOP mode)}: central values of polynomial coefficients $\vec p_{k}$ for each named polynomial $P_k$.
    \item \textbf{\texttt{observable\_central} (\emph{required} in SP mode; \emph{optional}  in FOP mode)}: central values of \emph{observable coefficients} $\vec o_{m}$, the polynomial coefficients for each observable $O_m$.
    \item \textbf{\texttt{observable\_uncertainties} (\emph{optional})}: uncertainties $\vec {\sigma}_{m}$ on the observable coefficients.
\end{itemize}
Polynomial coefficients are indexed by monomials, written as stringified tuples of model parameters, and complex parameters are handled through optional tagging of real and imaginary parts (see below for more details). Values are arrays of numbers defining predictions for a set of polynomials or observables (single-element arrays for a single polynomial or observable). The length of each array must match the number of entries in the corresponding \texttt{metadata.polynomial\_names} (for polynomials) and \texttt{metadata.observable\_names} (for observables). Missing monomials are implicitly treated as having zero coefficients.

%%%%%%%%%%%%%%%%%%%%%%%%%%%%%%%%%%%%%%%%%%%%%%%%%%%%%%%%%%%%%%%%%%%%%%%
%%%%%%%%%%%%%%%%%%%%%%%%%%%%%%%%%%%%%%%%%%%%%%%%%%%%%%%%%%%%%%%%%%%%%%%
%%%%%%%%%%%%%%%%%%%%%%%%%%%%%%%%%%%%%%%%%%%%%%%%%%%%%%%%%%%%%%%%%%%%%%%
\setminted[json]{
  frame=single,
  framesep=10pt,
  breaklines=true,      % wrap long lines
}
\begin{SpecSection}
\section{\texorpdfstring{Specification of Fields in the \texttt{POPxf}
\texttt{JSON}
Format}{Specification of Fields in the POPxf JSON Format}}\label{specification-of-fields-in-the-popxf-json-format}

A detailed specification of all fields in the \texttt{POPxf} data format
is given below. Each subsection describes the structure, expected data
type, and allowed values of the corresponding entries in the
\texttt{JSON} object. The format is divided into two main components:
the \texttt{metadata} and \texttt{data} fields. An additional
\texttt{\$schema} field is included to specify the version of the
\texttt{POPxf} \texttt{JSON} schema used. The data type \emph{object}
mentioned below refers to a \texttt{JSON} object literal and corresponds
to a set of key/value pairs representing named subfields. The data type
\emph{array} corresponds to an ordered list with elements of a specified
type. If the elements are arrays themselves, this corresponds to a
nested list or multi-dimensional array. If the elements are of any other
type, the array is a one-dimensional list. All quantities defined in
this specification refer to a single datafile. They may be indexed by a
superscript \((n)\) with \(n \in [1,N]\) to denote quantities in a
collection of \(N\) datafiles. This is particularly relevant for
discussing correlated predictions stored in separate files. Since this
specification focuses on the format of a single datafile, we will omit
the superscript \((n)\) to keep the notation concise. As a convention,
we assume that all dimensionful quantities are given in units of GeV.

\subsection{\texorpdfstring{\texttt{\$schema}
Field}{\$schema Field}}\label{schema-field}

The \texttt{\$schema} field allows identifying a \texttt{JSON} file as
conforming to the \texttt{POPxf} format and specifies the version of the
\texttt{POPxf} \texttt{JSON} schema used. It must be set to

\texttt{"https://json.schemastore.org/popxf-1.0.json"}

for files conforming to this version of the specification. The version
number will be incremented for future revisions of the \texttt{JSON}
schema.

\subsection{\texorpdfstring{\texttt{metadata}
Field}{metadata Field}}\label{metadata-field}

The \texttt{metadata} field contains all contextual and structural
information required to interpret the numerical predictions. It is a
\texttt{JSON} object with the following subfields:

\subsubsection{\texorpdfstring{\texttt{observable\_names} (required,
\emph{type: array of
string})}{observable\_names (required, type: array of string)}}\label{observable_names-required-type-array-of-string}

Array of \(M\) names identifying each observable \(O_m\). Must be an
array of unique, non-empty strings, with at least one entry.

Example:

\begin{minted}{json}
  "observable_names": ["observable1", "observable2", "observable3"]
\end{minted}

\subsubsection{\texorpdfstring{\texttt{parameters} (required,
\emph{type: array of
string})}{parameters (required, type: array of string)}}\label{parameters-required-type-array-of-string}

Array of \(S\) names identifying each model parameter \(C_s\) (e.g.,
Wilson coefficient names). Must be an array of unique, non-empty
strings, with at least one entry. In general, this includes
\(S_\mathbb{R}\) real-valued and \(S_\mathbb{C}\) complex-valued
parameters with \(S = S_\mathbb{R} + S_\mathbb{C}\). The real-valued
parameters and the real and imaginary parts of the complex-valued
parameters are used as the \(R=S_\mathbb{R} + 2\ S_\mathbb{C}\)
independent variables of all polynomial terms and can be grouped
together in a real-valued parameter vector \(\vec{C}\) of length \(R\).

Example:

\begin{minted}{json}
  "parameters": ["C1", "C2", "C3"]
\end{minted}

\subsubsection{\texorpdfstring{\texttt{basis} (required, \emph{type:
object})}{basis (required, type: object)}}\label{basis-required-type-object}

Defines the parameter basis (e.g.~an operator basis in an EFT). At least
one of the two subfields \texttt{wcxf} and \texttt{custom} has to be
present. If both subfields are present, any element of
\texttt{parameters} (see above) not belonging to the \texttt{wcxf} basis
is interpreted as belonging to the \texttt{custom} basis. The subfields
are defined as follows:

\begin{itemize}
\tightlist
\item
  \textbf{\texttt{wcxf} (optional, \emph{type: object})}: Specifies an
  EFT basis defined by the Wilson Coefficient exchange format (WCxf)~\cite{Aebischer:2017ugx}. This object contains the following fields:

  \begin{itemize}
  \tightlist
  \item
    \textbf{\texttt{eft} (required, \emph{type: string})}: EFT name
    defined by WCxf (e.g., \texttt{"SMEFT"})
  \item
    \textbf{\texttt{basis} (required, \emph{type: string})}: Operator
    basis name defined by WCxf (e.g., \texttt{"Warsaw"})
  \item
    \textbf{\texttt{sectors} (optional, \emph{type: array of string})}:
    Array of renormalisation-group-closed sectors of Wilson coefficients
    containing the Wilson coefficients given in \texttt{parameters} (see
    above). The available sectors for each EFT are defined by WCxf.
  \end{itemize}
\item
  \textbf{\texttt{custom} (optional, \emph{type: any})}: Field of any
  type and substructure to unambiguously specify any parameter basis not
  defined by WCxf.
\end{itemize}

Example:

\begin{minted}{json}
  "basis": {
    "wcxf": {
      "eft": "SMEFT",
      "basis": "Warsaw",
      "sectors": ["dB=de=dmu=dtau=0"]
    }
  }
\end{minted}

\subsubsection{\texorpdfstring{\texttt{polynomial\_names} (optional,
\emph{type: array of
string})}{polynomial\_names (optional, type: array of string)}}\label{polynomial_names-optional-type-array-of-string}

\emph{This field is required to express observables as functions of
polynomials. It requires the simultaneous presence of
\texttt{metadata.observable\_expressions} and
\texttt{data.polynomial\_central}.}

Array of \(K\) names identifying the individual polynomials \(P_k\) that
enter the observable predictions through the functions defined in
\texttt{metadata.observable\_expressions} (see below). Must contain
unique, non-empty strings.

Example:

\begin{minted}{json}
  "polynomial_names": ["polynomial 1", "polynomial 2"]
\end{minted}

\subsubsection{\texorpdfstring{\texttt{observable\_expressions}
(optional, \emph{type: array of
object})}{observable\_expressions (optional, type: array of object)}}\label{observable_expressions-optional-type-array-of-object}

\emph{This field is required to express observables as functions of
polynomials. It requires the simultaneous presence of
\texttt{metadata.polynomial\_names} and
\texttt{data.polynomial\_central}.}

Defines how each observable is constructed from the named polynomials.
Must be an array of \(M\) objects, one per observable. The length and
order of the array must match those of the \texttt{observable\_names}
field. Each object must contain:

\begin{itemize}
\tightlist
\item
  \textbf{\texttt{variables} (required, \emph{type: object})}: An object
  where each key is a string that is a Python-compatible variable name
  (used as variable in the \texttt{expression} field described below),
  and each value is a string identifying a polynomial name from
  \texttt{polynomial\_names}. For example,
  \texttt{\{"num":\ "polynomial\ 1",\ "den":\ "polynomial\ 2"\}}.
\item
  \textbf{\texttt{expression} (required, \emph{type: string})}: A
  Python-compatible mathematical expression using the variable names
  defined in \texttt{variables}, e.g.~\texttt{"num/den"}. The following
  elementary mathematical functions/constants may be used: \texttt{pi},
  \texttt{sqrt}, \texttt{sin}, \texttt{cos}, \texttt{tan},
  \texttt{arcsin}, \texttt{arccos}, \texttt{arctan}, \texttt{arctan2},
  \texttt{sinh}, \texttt{cosh}, \texttt{tanh}, \texttt{arcsinh},
  \texttt{arccosh}, \texttt{arctanh}, \texttt{exp}, \texttt{log}.
\end{itemize}

Example:

\begin{minted}{json}
  "observable_expressions": [
    {
      "variables": {
        "num": "polynomial 1",
        "den": "polynomial 2"
      },
      "expression": "num / den"
    },
    {
      "variables": {
        "num": "polynomial 2",
        "den": "polynomial 1"
      },
      "expression": "num / den"
    },
    {
      "variables": {
        "p1": "polynomial 1"
      },
      "expression": "sqrt(p1**2)"
    }
  ]
\end{minted}

\subsubsection{\texorpdfstring{\texttt{scale} (required, \emph{type:
number,
array})}{scale (required, type: number, array)}}\label{scale-required-type-number-array}

The renormalisation scale in GeV at which the parameter vector
\(\vec{C}\), the polynomial coefficients
\({\vec{p}_k \supset \vec{b}_k, \vec{c}_k, ...}\), and the observable
coefficients \({\vec{o}_m \supset \vec{b}_m, \vec{c}_m, ...}\) and their
uncertainties \(\vec{\sigma}_m\) are defined. The parameter vector
\(\vec{C}\) that enters a given polynomial \(P_k\) or observable \(O_m\)
has to be given at the same scale at which the polynomial coefficients
\(\vec{p}_k\) or observable coefficients \(\vec{o}_m\) are defined, such
that the polynomial or observable itself is scale-independent up to
higher-order corrections in perturbation theory. Scale dependence
cancels between the parameters \(\vec C\) and the coefficients
\(\vec{p}_k\) or \(\vec{o}_m\), and the constant terms \(a_k\) and
\(a_m\) are separately scale-independent.

This field can take one of two forms:

\begin{itemize}
\item
  \textbf{single number}: A common scale \(\mu\) at which all polynomial
  coefficients \(\vec p_k\) or observable coefficients \(\vec o_m\) are
  defined.

  \begin{itemize}
  \item
    If the observables \(O_m\) are expressed in terms of polynomials
    \(P_k\), the polynomials are functions of the parameters evolved to
    the common scale \(\mu\):
    \[P_k = a_{k} + \vec{C}(\mu) \cdot \vec{b}_{k}(\mu) + \dots\ \]
  \item
    If the observables \(O_m\) are themselves polynomials, they are
    themselves functions of the parameters evolved to the common scale
    \(\mu\): \[O_m = a_m + \vec{C}(\mu) \cdot \vec{b}_m(\mu) + \dots\ \]
  \end{itemize}
\item
  \textbf{array of numbers}: An array defining separate scales \(\mu_k\)
  of polynomial coefficients \(\vec p_k\) if
  \texttt{metadata.polynomial\_names} is present, or separate scales
  \(\mu_m\) of observable coefficients \(\vec o_m\) if
  \texttt{metadata.polynomial\_names} is absent.

  \begin{itemize}
  \item
    If \texttt{metadata.polynomial\_names} is present, the observables
    \(O_m\) are expressed in terms of polynomials \(P_k\) and each
    polynomial is a function of the parameters evolved to its
    corresponding scale \(\mu_k\):
    \[P_k = a_{k} + \vec{C}(\mu_k) \cdot \vec{b}_{k}(\mu_k) + \dots\ \]

    The length and order of the array defining the scales \(\mu_k\) must
    match those of the field \texttt{metadata.polynomial\_names}. To
    avoid ambiguities, the following restrictions apply to this case:

    \begin{itemize}
    \tightlist
    \item
      \texttt{data.observable\_central} must be absent;
    \item
      \texttt{data.observable\_uncertainties} must be absent or only
      define uncertainties for the parameter-independent terms
      (i.e.~only the SM uncertainties in EFT applications).
    \end{itemize}
  \item
    If \texttt{metadata.polynomial\_names} is absent, the observables
    \(O_m\) are themselves polynomials and each observable is a function
    of the parameters evolved to its corresponding scale \(\mu_m\):
    \[O_m = a_m + \vec{C}(\mu_m) \cdot \vec{b}_m(\mu_m) + \dots\ \]

    The length and order of the array defining the scales \(\mu_m\) must
    match those of the field \texttt{metadata.observable\_names}.
  \end{itemize}
\end{itemize}

Examples:

\begin{minted}{json}
  "scale": 91.1876
\end{minted}

\begin{minted}{json}
  "scale": [100.0, 200.0, 300.0, 400.0, 500.0]
\end{minted}

\subsubsection{\texorpdfstring{\texttt{polynomial\_degree} (optional,
\emph{type:
integer})}{polynomial\_degree (optional, type: integer)}}\label{polynomial_degree-optional-type-integer}

Specifies the maximum degree of polynomial terms included in the
expansion. If omitted, the default value is 2 (i.e., quadratic
polynomial). Values higher than 2 may be used to represent observables
involving higher-order terms in the model parameters. The current
implementation of the \texttt{JSON} schema defining the data format
supports values up to 5. Higher degrees are not prohibited in principle
but are currently unsupported to avoid excessively large data
structures.

Example:

\begin{minted}{json}
  "polynomial_degree": 2
\end{minted}

\subsubsection{\texorpdfstring{\texttt{reproducibility} (optional,
\emph{type: array of
object})}{reproducibility (optional, type: array of object)}}\label{reproducibility-optional-type-array-of-object}

Collects relevant data that may be required by a third party to
reproduce the prediction. Each element of the array should be an object
that corresponds to a step in the workflow and has three predefined
fields: \texttt{description}, \texttt{tool} and \texttt{inputs},
specified below. In addition, any additional fields containing data
deemed useful in this context can be included.

Schematic example:

\begin{minted}{json}
  "reproducibility": [
    {
      "description": "Description of the first step",
      "tool": { ... },
      "inputs": { ... }
    },
    {
      "description": "Description of the second step",
      "tool": { ... },
      "inputs": { ... }
    },
    ...
  ]
\end{minted}

The predefined fields are as follows:

\begin{itemize}
\item
  \textbf{\texttt{description} (optional, \emph{type: string})}:
  Free-form text description of the method and tool used in this step of
  obtaining the predictions.
\item
  \textbf{\texttt{inputs} (optional, \emph{type: object})}: Specifies
  the numerical values of input parameters used by the tool in producing
  the numerical values of the polynomial coefficients. Each entry maps
  an input name (a string) or a group of names (a stringified tuple such
  as
  \texttt{"(\textquotesingle{}m1\textquotesingle{},\textquotesingle{}m2\textquotesingle{})"})
  to one of the following:

  \begin{itemize}
  \tightlist
  \item
    A single number: interpreted as the central value of a single,
    uncorrelated input parameter without uncertainty;
  \item
    An object representing a uni- or multi-variate normal distribution
    describing one or more possibly correlated input parameters with
    uncertainties. This object can contain the subfields \texttt{mean},
    \texttt{std}, and \texttt{corr}. If the key of the object is a
    stringified tuple of \(N\) input names (e.g.,
    \texttt{"(\textquotesingle{}m1\textquotesingle{},\textquotesingle{}m2\textquotesingle{})"}
    with \(N = 2\)), describing a group of \(N\) possibly correlated
    input parameters, then \texttt{mean} and (if present) \texttt{std}
    must be arrays of length \(N\), and (if present) \texttt{corr} must
    be an \(N \times N\) matrix, expressed as an array of \(N\) arrays
    of \(N\) numbers. The subfields are defined as follows:

    \begin{itemize}
    \tightlist
    \item
      \textbf{\texttt{mean} (required, \emph{type: number, array})}:
      central value / mean; a single number for a single input name, or
      an array of numbers for a group of input names;
    \item
      \textbf{\texttt{std} (optional, \emph{type: number, array})}:
      uncertainty / standard deviation; a single number for a single
      input name, or an array of numbers for a group of input names;
    \item
      \textbf{\texttt{corr} (optional, \emph{type: array of array})}:
      correlation matrix; must only be used if a group of input names is
      given and requires the presence of \texttt{std}.
    \end{itemize}
  \item
    An object representing an arbitrary user-defined uni- or
    multi-variate probability distribution describing one or more input
    parameters. This object contains the following subfields:

    \begin{itemize}
    \tightlist
    \item
      \textbf{\texttt{distribution\_type} (required, \emph{type:
      string})}: a user-defined name identifying the probability
      distribution (e.g.~\texttt{"uniform"});
    \item
      \textbf{\texttt{distribution\_parameters} (required, \emph{type:
      object})}: an object where each key is a user-defined name of a
      parameter of the probability distribution, and each value is a
      single number in the univariate case, or an array of numbers or
      arrays in the multivariate case (e.g.~\texttt{\{"a":0,\ "b":1\}}
      for a uniform distribution with boundaries \(a\) and \(b\)).
    \item
      \textbf{\texttt{distribution\_description} (required, \emph{type:
      string})}: Description of the custom distribution implemented,
      defining the fields in \texttt{distribution\_parameters}.
    \end{itemize}
  \end{itemize}

  Example:

  In the example below, \texttt{"m1"} is an input parameter with no
  associated uncertainty, \texttt{"m2"} and \texttt{"m3"} are a pair of
  input parameters with correlated, Gaussian uncertainties, and
  \texttt{"m4"} is a parameter that is uniformly distributed between 0
  and 1.

  \begin{minted}{json}
    "inputs": {
      "m1": 1.0,
      "('m2','m3')": {
        "mean": [1.0, 2.0],
        "std": [0.1, 0.1],
        "corr": [
          [1.0, 0.3],
          [0.3, 1.0]
        ]
      },
      "m4": {
        "distribution_type": "uniform",
        "distribution_parameters": {
          "a": 0,
          "b": 1
        },
        "distribution_description": "Uniform distribution with boundaries $a$ and $b$."
      }
    }
  \end{minted}
\item
  \textbf{\texttt{tool} (optional, \emph{type: object})}: Provides
  free-form information about the tool, software or technique used in a
  particular step of the workflow. The predefined subfields are
  \texttt{name}, \texttt{version}, and \texttt{settings}. Any number of
  additional fields may be included to record or link to supplementary
  metadata, such as model information/configuration, perturbative order,
  scale choice, PDF sets, simulation settings, input parameter cards,
  etc. The predefined subfields are as follows:

  \begin{itemize}
  \tightlist
  \item
    \textbf{\texttt{name} (required, \emph{type: string})}: name of
    tool, e.g.~\texttt{"MadGraph5\_aMC@NLO"}, \texttt{"POWHEG"},
    \texttt{"SHERPA"}, \texttt{"WHIZARD"}, \texttt{"flavio"},
    \texttt{"FeynCalc"}, \texttt{"analytical\ calculation"}, \ldots{}
  \item
    \textbf{\texttt{version} (optional, \emph{type: string})}: version
    of the tool, e.g.~\texttt{"1.2"}
  \item
    \textbf{\texttt{settings} (optional, \emph{type: object})}: object
    containing information about the tool settings with free-form
    substructure. For example:

    \begin{itemize}
    \tightlist
    \item
      \texttt{perturbative\_order} (e.g.~\texttt{"LO"}, \texttt{"NLO"},
      \texttt{"NLOQCD"}, \ldots)
    \item
      \texttt{PDF}: name, version, and set of the PDF used.
    \item
      \texttt{UFO}: name and version of UFO model used, as well as any
      other relevant information such as flavor schemes or webpage link.
    \item
      \texttt{cuts}: Information about kinematical cuts specifying the
      phase space region over which the observable is computed
      (e.g.~acceptance effects, signal region definition, \ldots).
    \item
      \texttt{scale\_choice}: Nominal scale choice employed when
      computing the predictions. This could be an array of fixed scales
      or a string describing a dynamical scale choice like
      \texttt{"dynamical:HT/2"}. This field is particularly relevant
      when RGE effects are folded into the prediction, see the
      description of \texttt{metadata.scale} above.
    \item
      \texttt{renormalization\_scheme}: details of the renormalization
      scheme used in the computation.
    \item
      \texttt{covariant\_derivative\_sign}: sign convention used for the
      covariant derivative (\texttt{"+"} or \texttt{"-"}).
    \item
      \texttt{gamma5\_scheme}: scheme used for \(\gamma_5\) in
      dimensional regularization (\texttt{"BMHV"}, \texttt{"KKS"},
      \ldots).
    \item
      \texttt{evanescent}: details of the treatment of evanescent
      operators, e.g.~a reference to the scheme used.
    \item
      \texttt{approximations}: Any relevant approximations used, such as
      the use of the first leading-logarithmic approximation for RG
      evolution.
    \item
      any other relevant settings specific to the tool or calculation.
    \end{itemize}
  \end{itemize}

  Examples:

  \begin{minted}{json}
    "tool": {
      "name": "EFTTool",
      "version": "1.0.0"
    }
  \end{minted}

  \begin{minted}{json}
    "tool": {
      "name": "MadGraph5_aMC@NLO",
      "version": "3.6.2",
      "settings": {
        "UFO": {
          "name": "SMEFTUFO",
          "version": "1.0.0",
          "webpage": "https://smeftufo.io"
        },
        "PDF": {
          "name": "LHAPDF",
          "version": "6.5.5",
          "set": "331700"
        },
        "perturbative_order": "NLOQCD",
        "scale_choice": [91.1876, 125.0]
      }
    }
  \end{minted}

  \begin{minted}{json}
    "tool": {
      "name": "AnalysisTool",
      "version": "1.0.0",
      "settings": {
        "cuts": {
          "pT_min": 20.0,
          "eta_max": 2.5
        },
        "code": "https://coderepository.com/analysis/example"
      }
    }
  \end{minted}

  \begin{minted}{json}
    "tool": {
      "name": "analytical calculation",
      "settings": {
        "gamma5_scheme": "KKS",
        "covariant_derivative_sign": "-",
        "renormalization_scheme": "MSbar (WCs), On-shell (mass, aS, aEW)",
        "evanescent": "https://doi.org/10.1016/0550-3213(90)90223-Z"
      }
    }
  \end{minted}

  \begin{minted}{json}
    "tool": {
      "name": "RGEtool",
      "version": "1.0.0",
      "settings": {
        "perturbative_order": "one-loop",
        "method": "evolution matrix formalism"
      }
    }
  \end{minted}
\end{itemize}

\subsubsection{\texorpdfstring{\texttt{misc} (optional, \emph{type:
object})}{misc (optional, type: object)}}\label{misc-optional-type-object}

Optional free-form metadata for documentation purposes. May include
fields such as authorship, contact information, date, description of the
observable, information identifying the associated correlation file
(e.g.~hash value or filename), or external references. The format is
unrestricted, allowing any \texttt{JSON}-encodable content.

Example:

\begin{minted}{json}
  "misc": {
    "author": "John Doe",
    "contact": "john.doe@example.com",
    "description": "Example dataset",
    "URL": "johndoe.com/exampledata",
    "correlation_file": "correlations.json",
    "correlation_file_hash": "AB47BG3F11DA7DCAA5726008BAAFE176"
  }
\end{minted}

\subsection{\texorpdfstring{\texttt{data}
Field}{data Field}}\label{data-field}

The \texttt{data} field contains the numerical representation of all
polynomial terms, which define the polynomials \(P_k\) and observables
\(O_m\). This information is provided in terms of central values of
polynomial coefficients \(\vec{p}_k\) and observable coefficients
\(\vec{o}_m\), and uncertainties of observable coefficients
\(\vec{\sigma}_m\).

Each component of \(\vec{o}_m\), \(\vec{p}_k\), and \(\vec{\sigma}_m\)
is labelled by a \emph{monomial key}, written as a stringified tuple of
model parameters (e.g., Wilson coefficients) defined in
\texttt{metadata.parameters}. For example, the key
\texttt{"(\textquotesingle{}C1\textquotesingle{},\ \textquotesingle{}C2\textquotesingle{})"}
corresponds to the monomial \(C_1 C_2\). While the model parameters can
be complex numbers, the monomials are defined for the real and imaginary
parts of the model parameters (see below) and are therefore strictly
real. The format and conventions for monomial keys are as follows:

\begin{itemize}
\tightlist
\item
  Each key is a string representation of a Python-style tuple: a
  comma-separated array of strings enclosed in parentheses.
\item
  The length of the tuple is determined by the polynomial degree \(d\),
  as defined by the \texttt{metadata} field \texttt{polynomial\_degree}
  (default value: \(d=2\), i.e.~quadratic polynomial, if
  \texttt{polynomial\_degree} is omitted). The tuple length equals
  \(d\), unless a real/imaginary tag is included (see below), in which
  case the length is \(d+1\).
\item
  The first \(d\) entries in the tuple are model parameter names, as
  defined in the \texttt{metadata} field \texttt{parameters}. These
  names must be sorted alphabetically to ensure unique monomial keys
  (assuming the same sorting rules as Python's \texttt{sort()} method
  which sorts alphabetically according to ASCII or UNICODE-value, where
  all upper-case letters come before all lower-case letters, and shorter
  strings take precedence). Empty strings
  \texttt{\textquotesingle{}\textquotesingle{}} are used to represent
  constant terms (equivalent to \(1\)) and to pad monomials of lower
  degree. For example, for a quadratic polynomial in real parameters
  (see below for how complex parameters are handled):

  \begin{itemize}
  \tightlist
  \item
    A constant \(1\) is written as
    \texttt{"(\textquotesingle{}\textquotesingle{},\textquotesingle{}\textquotesingle{})"},
  \item
    A linear term \(C_1\) is written as
    \texttt{"(\textquotesingle{}\textquotesingle{},\ \textquotesingle{}C1\textquotesingle{})"},
  \item
    A quadratic term \(C_1 C_2\) is written as
    \texttt{"(\textquotesingle{}C1\textquotesingle{},\ \textquotesingle{}C2\textquotesingle{})"}.
  \end{itemize}
\item
  To handle complex parameters, the tuple may optionally include a
  real/imaginary tag as its final element. This tag consists of
  \texttt{R} (real) and \texttt{I} (imaginary) characters, and its
  length must match the polynomial degree \(d\). It indicates whether
  each parameter refers to its real or imaginary part. For example:

  \begin{itemize}
  \tightlist
  \item
    \texttt{"(\textquotesingle{}\textquotesingle{},\ \textquotesingle{}C1\textquotesingle{},\ \textquotesingle{}RI\textquotesingle{})"}
    corresponds to \(\mathrm{Im}(C_1)\);
  \item
    \texttt{"(\textquotesingle{}C1\textquotesingle{},\ \textquotesingle{}C2\textquotesingle{},\ \textquotesingle{}IR\textquotesingle{})"}
    corresponds to \(\mathrm{Im}(C_1)\mathrm{Re}(C_2)\).
  \end{itemize}
\item
  If the real/imaginary tag is omitted, the parameters are assumed to be
  real. For example:

  \begin{itemize}
  \tightlist
  \item
    \texttt{"(\textquotesingle{}\textquotesingle{},\ \textquotesingle{}C1\textquotesingle{})"}
    corresponds to \(\mathrm{Re}(C_1)\);
  \item
    \texttt{"(\textquotesingle{}C1\textquotesingle{},\ \textquotesingle{}C2\textquotesingle{})"}
    corresponds to \(\mathrm{Re}(C_1)\mathrm{Re}(C_2)\).
  \end{itemize}
\end{itemize}

These conventions ensure a canonical and unambiguous representation of
polynomial terms while offering flexibility in the naming of model
parameters. Missing monomials are implicitly treated as having zero
coefficients.

The \texttt{data} field is a \texttt{JSON} object with the following
subfields:

\subsubsection{\texorpdfstring{\texttt{polynomial\_central} (optional,
\emph{type:
object})}{polynomial\_central (optional, type: object)}}\label{polynomial_central-optional-type-object}

\emph{This field is required to express observables as functions of
polynomials. It requires the simultaneous presence of
\texttt{metadata.polynomial\_names} and
\texttt{metadata.observable\_expressions}.}

An object representing the central values of the polynomial coefficients
\(\vec{p}_k\) for each named polynomial \(P_k\). Each key must be a
monomial key as defined above. Each value must be an array of \(K\)
numbers whose order matches \texttt{metadata.polynomial\_names}.

Example:

Specifying two polynomials, \(P_k\), given in terms of two complex
parameters \(C_1\) and \(C_2\) as \[
\begin{aligned}
    P_1 &= 1.0 + 1.2 \ \mathrm{Im}(C_1) + 0.8 \ \mathrm{Re}(C_1) \mathrm{Re}(C_2) + 0.5 \ \mathrm{Re}(C_1) \mathrm{Im}(C_2)+ 0.2 \ \mathrm{Im}(C_1) \mathrm{Im}(C_2)\ , \\
    P_2 &= 1.1 + 1.3 \ \mathrm{Im}(C_1)  + 0.85 \ \mathrm{Re}(C_1) \mathrm{Re}(C_2) + 0.55 \ \mathrm{Re}(C_1) \mathrm{Im}(C_2)+ 0.25 \ \mathrm{Im}(C_1) \mathrm{Im}(C_2)\ .
\end{aligned}
\]

\begin{minted}{json}
  "polynomial_central": {
    "('', '', 'RR')": [1.0, 1.1],
    "('', 'C1', 'RI')": [1.2, 1.3],
    "('C1', 'C2', 'RR')": [0.8, 0.85],
    "('C1', 'C2', 'RI')": [0.5, 0.55],
    "('C1', 'C2', 'II')": [0.2, 0.25]
  }
\end{minted}

\subsubsection{\texorpdfstring{\texttt{observable\_central} (optional,
\emph{type:
object})}{observable\_central (optional, type: object)}}\label{observable_central-optional-type-object}

An object representing the central values of the observable coefficients
\(\vec{o}_m\) for each observable{~}\(O_m\). In case the observables are
not themselves polynomials, the observable coefficients correspond to
the polynomial approximation of the observables obtained from a Taylor
expansion of the observable expressions defined in
\texttt{metadata.observable\_expressions}. Each key must be a monomial
key as defined above. Each value must be an array of \(M\) numbers whose
order matches \texttt{metadata.observable\_names}.

Example:

Specifying three observable predictions, \(O_{m}\), given in terms of
the three real parameters \(C_1\), \(C_2\), and \(C_3\) as \[
\begin{aligned}
    O_1 &= 1.0 + 1.2 \ C_1 + 1.4 \ C_1C_2+ 1.6 \ C_1C_3\ , \\
    O_2 &= 1.1 + 1.3 \ C_1 + 1.5 \ C_1C_2+ 1.7 \ C_1C_3\ , \\
    O_3 &= 2.3 + 0.3\ C_1 + 0.7 \ C_1C_2 + 0.5 \ C_1C_3\ .
\end{aligned}
\]

\begin{minted}{json}
  "observable_central": {
    "('', '')": [1.0, 1.1, 2.3],
    "('', 'C1')": [1.2, 1.3, 0.3],
    "('C1', 'C2')": [1.4, 1.5, 0.7],
    "('C1', 'C3')": [1.6, 1.7, 0.5]
  }
\end{minted}

\subsubsection{\texorpdfstring{\texttt{observable\_uncertainties}
(optional, \emph{type:
object})}{observable\_uncertainties (optional, type: object)}}\label{observable_uncertainties-optional-type-object}

An object representing the uncertainties on the observable coefficients
\(\vec{\sigma}_m\) for each observable{~}\(O_m\). In case the
observables are not themselves polynomials, the observable coefficients
correspond to the polynomial approximation of the observables obtained
from a Taylor expansion of the observable expressions defined in
\texttt{metadata.observable\_expressions}. The fields specify the nature
of quoted uncertainty. In many cases there may only be a single
top-level field, \texttt{"total"}, but multiple fields can be used to
specify a breakdown into several sources of uncertainty (e.g.,
statistical, scale, PDF, \ldots). To avoid mistakes, the names of the
top-level fields must not have the format of a monomial key (i.e.,
stringified tuples as defined above). The value of each top-level field
can either be an object or an array of floats. Objects must have the
same structure as \texttt{observable\_central}, arrays must have length
\(M\). If instead of an object, an array of floats is specified, it is
assumed to correspond to the parameter independent uncertainty only
(e.g.~the uncertainty on the SM prediction). This would be equivalent to
specifying an object containing a single element with the monomial key
of the constant term
(e.g.~\texttt{"(\textquotesingle{}\textquotesingle{},\textquotesingle{}\textquotesingle{})"}
for a quadratic polynomial).

Examples:

\begin{minted}{json}
  "observable_uncertainties": {
    "total": {
      "('', '')": [0.05, 0.06, 0.01],
      "('', 'C1')": [0.1, 0.12, 0.01],
      "('C1', 'C2')": [0.02, 0.03, 0.02],
      "('C1', 'C3')": [0.05, 0.06, 0.01]
    }
  }
\end{minted}

Specifying only the SM uncertainties:

\begin{minted}{json}
  "observable_uncertainties": {
    "total": [0.05, 0.06, 0.01]
  }
\end{minted}

Specifying an uncertainty breakdown:

\begin{minted}{json}
  "observable_uncertainties": {
    "MC_stats": {
      "('', '')": [0.002, 0.0012, 0.001],
      "('', 'C1')": [0.001, 0.0015, 0.0001]
    },
    "scale": {
      "('', '')": [0.04, 0.05, 0.06],
      "('', 'C1')": [0.1, 0.12, 0.01]
    },
    "PDF": {
      "('', '')": [0.03, 0.04, 0.05],
      "('', 'C1')": [0.02, 0.08, 0.01]
    }
  }
\end{minted}
\pagebreak
Specifying a breakdown for SM uncertainties only:

\begin{minted}{json}
  "observable_uncertainties": {
    "MC_stats": [0.002, 0.0012, 0.001],
    "scale": [0.04, 0.05, 0.06],
    "PDF": [0.03, 0.04, 0.05]
  }
\end{minted}

\end{SpecSection}
%%%%%%%%%%%%%%%%%%%%%%%%%%%%%%%%%%%%%%%%%%%%%%%%%%%%%%%%%%%%%%%%%%%%%%%
%%%%%%%%%%%%%%%%%%%%%%%%%%%%%%%%%%%%%%%%%%%%%%%%%%%%%%%%%%%%%%%%%%%%%%%
%%%%%%%%%%%%%%%%%%%%%%%%%%%%%%%%%%%%%%%%%%%%%%%%%%%%%%%%%%%%%%%%%%%%%%%

\section{Data structure for correlations of observables}\label{sec:correlation_files}

The \texttt{JSON} structure defined in~\cref{sec:overview,specification-of-fields-in-the-popxf-json-format} contains the main information on observable predictions including their central values and uncertainties. Since observables in one \texttt{POPxf} data file (labelled by index $n$) can be correlated with observables in another \texttt{POPxf} data file (labelled by index $n^\prime$), we define correlations in a separate data structure that is implemented in a separate \texttt{POPxf} correlation file.

\subsection{File formats}

We support two different file formats for \texttt{POPxf} correlation data:
\begin{itemize}
 \item \texttt{JSON} files are primarily used for parameter-independent correlations or correlations between a relatively small number of observables. In these cases, correlation matrices typically contain $\mathcal{O}(10)$-$\mathcal{O}(1000)$ rows and columns -- a size for which the \texttt{JSON} data format is still suitable.
 \item \texttt{HDF5} files are primarily used for parameter-dependent correlations or correlations between a large number of observables. In these cases, correlation matrices can have $\mathcal{O}(10^4)$-$\mathcal{O}(10^6)$ rows and columns -- a size for which the Hierarchical Data Format version 5 (\texttt{HDF5}) is particularly well suited, as it is specifically designed for storing large numerical arrays.
\end{itemize}
For both file formats, we encode the correlation matrices in the same hierarchical data structure, with differences only in syntax: in \texttt{JSON}, the data is represented by \texttt{JSON} \emph{objects} containing \emph{key-value pairs with nested arrays as values}, whereas in \texttt{HDF5}, the same structure is realized through \texttt{HDF5} \emph{groups} containing \texttt{HDF5} \emph{datasets}.
In order to use a common language for both \texttt{JSON} and \texttt{HDF5} data structures, we will use the \texttt{HDF5} terms for the correpsonding \texttt{JSON} structures: we denote a \emph{key-value pair where the value is a \texttt{JSON} object} as \emph{group} and a \emph{key-value pair where the value is a (nested) array} as \emph{dataset}. Instead of \emph{key} and \emph{value}, we will use \emph{name} and \emph{content} (of a group or dataset).
The mapping between \texttt{JSON} and \texttt{HDF5} data structures and the common language used in the following are shown in \cref{tab:JSON-HDF5}.

\begin{table}[h]
\setcellgapes{4pt}
\makegapedcells
\center
\begin{tabular}{ccccc}
 \texttt{JSON} && \texttt{HDF5} && common language\\
 \hline
 \makecell{key-value pair\\with \texttt{JSON} object as value} & $\leftrightarrow$ & \texttt{HDF5} group & $\leftrightarrow$ & group \\
 \makecell{key-value pair\\with (nested) array as value} & $\leftrightarrow$ & \texttt{HDF5} dataset & $\leftrightarrow$ & dataset \\
 key & $\leftrightarrow$ & name of \texttt{HDF5} group or dataset & $\leftrightarrow$ & name\\
 value  & $\leftrightarrow$ & content of \texttt{HDF5} group or dataset & $\leftrightarrow$ & content\\
\end{tabular}
\caption{Mapping between \texttt{JSON} and \texttt{HDF5} data structures and the common language used for describing \texttt{POPxf} correlation data.}\label{tab:JSON-HDF5}
\end{table}

\subsection{File structure}

To encode correlation matrices $
 [\rho_{m m^\prime}^{(n n^\prime)}]_{\alpha{\alpha^\prime}}$ (as defined in~\cref{sec:uncertainties}) in a \texttt{POPxf} correlation file, we define one top-level entry for each pair of correlated \texttt{POPxf} data files, i.e.\ for each combination of $(n n^\prime)$ indices with $n\geq n^\prime$. The correlations for $n < n^\prime$ are related to those for $n > n^\prime$ by $
 [\rho_{m m^\prime}^{(n n^\prime)}]_{\alpha{\alpha^\prime}}=
 [\rho_{m^\prime m}^{(n^\prime n)}]_{{\alpha^\prime}\alpha}$.
 Each top-level entry in the \texttt{POPxf}  correlation file is a \emph{group} with unique name.
 The names are arbitrary, but for efficient lookup we recommend using specific hash values as names, which we describe in \cref{sec:corr_hash_vals}.

\texttt{POPxf} correlation files in  \texttt{JSON} format contain an additional top-level field \texttt{\$schema}, which declares the  \texttt{JSON} schema specification used by the file. This allows identifying a given  \texttt{JSON} file as a \texttt{POPxf} correlation file and validating its structure. The value must be

\texttt{https://json.schemastore.org/popxf-corr-1.0.json}

for the first version of this correlation format and the version number will be incremented if the format is extended in the future. \texttt{POPxf} correlation files in \texttt{HDF5} format contain the same value as an \texttt{HDF5} top-level \emph{attribute} called \texttt{\$schema}. This allows converting \texttt{POPxf} correlation files from \texttt{HDF5} to \texttt{JSON} and vice versa.

\begin{figure}
\begin{minipage}{0.47\textwidth}
\setminted{highlightcolor=white}
\texttt{JSON} file
\begin{minted}[frame=single,framesep=10pt,escapeinside=@@]{json}
{
  "$schema": "https://...",
  "<name of first entry>": {
    "row_names": [...],
    "col_names": [...],
    "correlations": {
      "<source 1>":[[...],...],
      "<source 2>":[[...],...],
      ...
    }
  },
  "<name of second entry>": {
    "row_names": [...],
    "col_names": [...],
    "correlations": {
      "<source 1>":[[...],...],
      "<source 2>":[[...],...],
      ...
    }
  },
  ...
}
\end{minted}
\end{minipage}
\hskip0.02\textwidth
\begin{minipage}{0.5\textwidth}
\renewcommand*\DTstyle{\normalfont}
\setlength{\DTbaselineskip}{17pt}
\DTsetlength{0.2em}{0.5em}{0.2em}{0.4pt}{1.6pt}
\dirtree{%
.1 \texttt{HDF5} file.
.2 attribute: \texttt{\$schema} (\emph{type: string}).
.2 group: \texttt{<name of first entry>}.
.3 dataset: \texttt{row\_names} (\emph{type: string}).
.3 dataset: \texttt{col\_names} (\emph{type: string}).
.3 group: \texttt{correlations}.
.4 dataset: \texttt{<source 1>} (\emph{type: numeric}).
.5 attribute: \texttt{scale\_factor} (\emph{optional}).
.4 dataset: \texttt{<source 2>} (\emph{type: numeric}).
.5 attribute: \texttt{scale\_factor} (\emph{optional}).
.4 \dots.
.2 group: \texttt{<name of second entry>}.
.3 dataset: \texttt{row\_names} (\emph{type: string}).
.3 dataset: \texttt{col\_names} (\emph{type: string}).
.3 group: \texttt{correlations}.
.4 dataset: \texttt{<source 1>} (\emph{type: numeric}).
.5 attribute: \texttt{scale\_factor} (\emph{optional}).
.4 dataset: \texttt{<source 2>} (\emph{type: numeric}).
.5 attribute: \texttt{scale\_factor} (\emph{optional}).
.4 \dots.
.2 \dots.
}
\end{minipage}
\caption{Structure of \texttt{POPxf} correlation file in \texttt{JSON} (left) and \texttt{HDF5} (right).}\label{fig:corr}
\end{figure}

 \subsubsection{Structure of top-level entries}
Each top-level entry contains three elements:
\begin{itemize}
\item \textbf{\texttt{row\_names} (\emph{dataset, type: string})}: array of observable names equal to the field \linebreak \texttt{metadata.observable\_names} in data file~$n$. These observable names correspond to the row index~$m$ of $[\rho_{m m^\prime}^{(n n^\prime)}]_{\alpha{\alpha^\prime}}$.
 \item \textbf{\texttt{col\_names} (\emph{dataset, type: string})}: array of observable names equal to the field \linebreak \texttt{metadata.observable\_names} in data file~$n^\prime$. These observable names correspond to the column index~$m^\prime$.
  \item
  \textbf{\texttt{correlations} (\emph{group})}: this group contains one numeric \emph{dataset} for each source of correlated uncertainty (e.g., MC statistics, scale, PDF, etc.).
  The names of the datasets have to match the keys of the correlated \texttt{data.observable\_uncertainties} fields in data files $n$ and $n^\prime$.
  If uncertainties are not broken down into several sources, only a single dataset named \texttt{total} may be present.
  Keys in the \texttt{data.observable\_uncertainties} fields of data files $n$ and $n^\prime$ for which no dataset is included, and keys that are only present in one of these files are assumed to correspond to uncorrelated uncertainties.
  The content of each included dataset is a multidimensional numerical array (nested array in \texttt{JSON}) containing the correlation coefficients.
 The shape of this array depends on whether the uncertainties and correlations are parameter-independent or parameter-dependent:
 \begin{itemize}
  \item For parameter-independent correlations, the array has two dimensions:
  \begin{enumerate}
      \item axis corresponding to the observable names in \texttt{row\_names}, i.e.\ those in \linebreak \texttt{metadata.observable\_names} in data file~$n$. The number and order of the entries has to match those of the observable names in \texttt{row\_names}.
      \item axis corresponding to the observable names in \texttt{col\_names}, i.e.\ those in \linebreak \texttt{metadata.observable\_names} in data file~$n^\prime$. The number and order of the entries has to match those of the observable names in \texttt{col\_names}.
  \end{enumerate}
  \item For parameter-dependent correlations, the array has four dimensions:
  \begin{enumerate}
      \item axis corresponding to the observable names in \texttt{row\_names}, i.e.\ those in \linebreak \texttt{metadata.observable\_names} in data file~$n$. The number and order of the entries has to match those of the observable names in \texttt{row\_names}.
      \item axis corresponding to the observable names in \texttt{col\_names}, i.e.\ those in \linebreak \texttt{metadata.observable\_names} in data file~$n^\prime$. The number and order of the entries has to match those of the observable names in \texttt{col\_names}.
      \item axis corresponding to the parameter monomials used as keys in the field \linebreak \texttt{data.observable\_central} in data file $n$. The number and order of the entries has to match those of the alphabetically sorted monomial keys.
      \item axis corresponding to the parameter monomials used as keys in the field \linebreak \texttt{data.observable\_central} in data file $n^\prime$. The number and order of the entries has to match those of the alphabetically sorted monomial keys.
  \end{enumerate}
  In \texttt{HDF5} correlation files, each dataset within a given \texttt{correlations} element has an optional \emph{attribute} \texttt{scale\_factor}, which contains a single floating point number. If present, all numerical values in the dataset have to be multiplied by \texttt{scale\_factor} to obtain the correlation coefficients. The main use-case for the \texttt{scale\_factor} attribute is to reduce the file size by implementing the correlation coefficients in a numerical format other than floating point numbers between $-1.0$ and $+1.0$. For example, by setting \texttt{scale\_factor} to $1/32767$, 16-bit signed integers between $-32767$ and $+32767$ can be used to represent correlation coefficients. If absent, \texttt{scale\_factor} is assumed to be $1.0$. \texttt{scale\_factor} is not supported in \texttt{JSON} files. If file size matters, \texttt{HDF5} should be used instead of \texttt{JSON}.
 \end{itemize}
\end{itemize}
For each pair of correlated \texttt{POPxf} data files, the corresponding correlation data can be identified by matching the \texttt{observable\_names} metadata fields of these files to the \texttt{row\_names} and \texttt{col\_names} elements in a \texttt{POPxf} correlation file.
For a more efficient lookup, we recommend creating hash values from the \texttt{row\_names} and \texttt{col\_names} elements and using them as the names of the top-level entries, as described in the following section.
The file structure in both file formats, \texttt{JSON} and \texttt{HDF5}, is illustrated in \cref{fig:corr}.

\subsubsection{Hash values for efficient lookup of top-level entries}\label{sec:corr_hash_vals}

For each pair of correlated \texttt{POPxf} data files, the corresponding correlation data can, in principle, be located by loading all top-level entries in a \texttt{POPxf} correlation file and comparing the \texttt{row\_names} and \texttt{col\_names} elements to the \texttt{observable\_names} metadata fields of the given pair of \texttt{POPxf} data files.
However, this procedure requires reading the content of all \texttt{row\_names} and \texttt{col\_names} elements and is therefore inefficient.
To enable faster lookup, we recommend computing hash values from \texttt{row\_names} and \texttt{col\_names} and using them as the names of the corresponding top-level entries in the \texttt{POPxf} correlation file.
Then, to retrieve the correlation data for a given pair of correlated \texttt{POPxf} data files, it is sufficient to compute the same hash value from the \texttt{observable\_names} metadata fields of these files, which can then be used to directly access the matching entry in the \texttt{POPxf} correlation file.

To standardise the computation of hash values, we adopt the following procedure for a given pair of \texttt{row\_names} and \texttt{col\_names}:
\begin{itemize}
 \item Join all observable names in the arrays \texttt{row\_names} and \texttt{col\_names} into a single string each, using the vertical bar character ``\texttt{|}'' as the separator.
   Before joining, escape all existing vertical bar characters within observable names using a backslash ``\texttt{\textbackslash}'', and escape any existing backslashes by doubling them.
 \item Concatenate the two resulting strings (from \texttt{row\_names} and \texttt{col\_names}) using two vertical bars ``\texttt{||}'' as the separator to form a single combined string.
 \item Compute the MD5 message digest of the UTF-8 encoded string and represent it as a 128-bit hexadecimal value.
   The MD5 algorithm is chosen for its simplicity and wide availability.
\end{itemize}
The resulting hash value uniquely identifies a pair of \texttt{row\_names} and \texttt{col\_names}.

For a given pair of correlated \texttt{POPxf} data files, it is not known \emph{a priori} which of them corresponds to the rows or to the columns.
By computing both possible hash values and checking which one is present in the correlation file, one can identify the corresponding correlation data and, at the same time, determine which file corresponds to the rows and which to the columns.
Using these hash values as top-level entry names ensures unambiguous identification and enables efficient access to the correlation data without requiring a full scan of the file contents.

%%%%%%%%%%%%%%%%%%%%%%%%%%%%%%%%%%%%%%%%%%%%%%%
\section{Conclusions and Outlook}
\label{sec:conclusions}
In this note, we have proposed the Polynomial Observable Prediction Exchange Format, \texttt{POPxf}, a standardised, machine-readable data format for sharing theoretical predictions that can be expressed as (functions of) polynomials in the model parameters. While the focus of this format is in EFT applications, it remains general enough to apply to other related cases. \texttt{POPxf} allows for the encoding of observable predictions via their monomial coefficients, defining observables as functions of polynomials, and for the specification of theoretical uncertainties and their possibly parameter-dependent correlations. The format allows for the inclusion of ample metadata, to maximise the reproducibility of published predictions. Adopting a common standard will reduce the duplication of efforts, facilitate validation, and maximise the accessibility and impact of such theoretical predictions.

The concrete specification of the \texttt{POPxf} data format in terms of \texttt{JSON} schemas is hosted in a public repository at \url{https://github.com/pop-xf}, which also includes a lightweight validator and associated command line tool as well as a collection of example files.

\acknowledgments
 This work was done on behalf of the LHC Higgs and EFT Working Groups and we would like to
thank all members for stimulating discussions that led to this document. We would also like to thank the CERN Theory division for their hospitality while the format was being developed. K.M. is supported by an Ernest Rutherford Fellowship from the STFC, Grant No. ST/X004155/1. A.S. is supported by the Slovenian Research and Innovation Agency (Grant No. J1-50219 and research core funding No. P1-0035).

\renewcommand{\theHsection}{A\arabic{section}}
\appendix
\section{Examples of complete \texttt{JSON} structures}\label{app:examples}
We give a few simplified examples of complete \texttt{JSON} files in the \texttt{POPxf} format below. The full examples including additional parameter dependencies and reproducibility information can be found at \url{https://github.com/pop-xf/examples}.

\subsection{Single polynomial mode}
The SP mode example below implements a SMEFT prediction for the partial $W$-boson width to a muon and muon-antineutrino, $\Gamma(W^-\to\mu^-\bar{\nu}_\mu)$, in the $(m_W, G_F,m_Z)$ input scheme.  The observable depends on 3 coefficients in the Warsaw basis~\cite{Grzadkowski:2010es} as implemented in the \texttt{SMEFTatNLO} UFO model~\cite{Degrande:2020evl} which was used to compute the predictions with \texttt{MadGraph5\_aMC@NLO}~\cite{Alwall:2014hca}, as indicated by the \texttt{"custom"} field of \texttt{metadata.basis} and the \texttt{"tool"} field of the entry in \texttt{metadata.reproducibility}.
\pagebreak
\begin{minted}[frame=single,framesep=9pt,fontsize=\footnotesize]{json}
{
  "$schema": "https://json.schemastore.org/popxf-1.0.json",
  "metadata": {
    "basis": {
      "custom": {
        "eft": "SMEFT",
        "basis": "SMEFTatNLO",
        "definition": "https://feynrules.irmp.ucl.ac.be/wiki/SMEFTatNLO"
      }
    },
    "scale": 80.387,
    "parameters": [ "c3pl1", "c3pl2", "cll" ],
    "observable_names": ["Gamma(W -> mu nu_m)"],
    "reproducibility": [{
      "description": "Fixed-order Monte Carlo computation.",
      "tool": {
        "name": "MadGraph5_aMC@NLO",
        "version": "3.4.1",
        "settings":{ "UFO":"SMEFTatNLO 1.0.2",  "perturbative_order":"LO" }
      },
      "inputs":{ "m_W": 80.387, "G_F": 1.1663787e-5, "m_Z": 91.1876, "Lambda":1000 }
    }],
    "misc":{
      "description":"W-boson partial width to mu nu_m [GeV], (m_W,G_F,m_Z) scheme",
      "author": ["E. Celada", "L. Mantani", "K. Mimasu"],
      "contact": "ken.mimasu@soton.ac.uk"
    }
  },
  "data": {
    "observable_central": {
       "('', '')": [0.22729],
       "('', 'c3pl1')": [-0.0137796],
       "('', 'c3pl2')": [0.0137786],
       "('', 'cll')": [0.0137796],
       "('c3pl1', 'c3pl1')": [0.000208845],
       "('c3pl2', 'c3pl2')": [0.00020885],
       "('cll', 'cll')": [0.00020884],
       "('c3pl1', 'c3pl2')": [-0.00041769],
       "('c3pl1', 'cll')": [-0.00041768],
       "('c3pl2', 'cll')": [0.00041768]
    }
  }
}
\end{minted}
% \newpage
\subsection{Function-of-polynomials mode}
The FOP mode example below implements a related SMEFT prediction for the ratios of partial $W$-boson widths to different lepton generations, $R_{\ell_i\ell_j}\!=\Gamma(W^-\to\ell_i^-\bar{\nu}_{\ell_i})/\Gamma(W^-\to\ell_j^-\bar{\nu}_{\ell_j})$, this time in the $(\alpha_{EW}, G_F,m_Z)$ input scheme, computed using \texttt{flavio}~\cite{Straub:2018kue}. The input polynomials encode the dependence on the partial widths and the observable expressions encode their ratios. For brevity, the dependence is encoded for a subset of three Wilson coefficients in the Warsaw basis as implemented in the corresponding \texttt{WCxf} definition, and information about input parameters in the \texttt{reproducibility} field is omitted. The complete example file containing the full dependence on all the contributing Wilson coefficients and including the entire \texttt{reproducibility} field can be found at \url{https://github.com/pop-xf/examples} under the name \texttt{Wlnu.json}.
\begin{minted}[frame=single,framesep=9pt,fontsize=\footnotesize]{json}
{
  "$schema": "https://json.schemastore.org/popxf-1.0.json",
  "metadata": {
    "observable_names": ["Rmue(W->lnu)", "Rtaue(W->lnu)", "Rtaumu(W->lnu)"],
    "parameters": ["phil3_11", "phil3_22", "phil3_33"],
    "basis": {
      "wcxf": {
        "eft": "SMEFT",
        "basis": "Warsaw",
        "sectors": ["dB=de=dmu=dtau=0"]
      }
    },
    "polynomial_names": ["Gamma(W->enu)", "Gamma(W->munu)", "Gamma(W->taunu)"],
    "observable_expressions": [
      {
        "expression": "num / den",
        "variables": {"num": "Gamma(W->munu)", "den": "Gamma(W->enu)"}
      },
      {
        "expression": "num / den",
        "variables": {"num": "Gamma(W->taunu)", "den": "Gamma(W->enu)"}
      },
      {
        "expression": "num / den",
        "variables": {"num": "Gamma(W->taunu)", "den": "Gamma(W->munu)"}
      }
    ],
    "scale": 80.387,
    "reproducibility": [{"tool": {"name": "flavio", "version": "2.6.2"}}],
    "misc": {
      "description": "Using the (alpha, G_F, m_Z) input scheme.",
      "author": ["A. Smolkovic", "P. Stangl"]
    }
  },
  "data": {
    "polynomial_central": {
      "('', '')": [0.227, 0.227, 0.227],
      "('', 'phil3_11')": [7737.419, -19812.903, -19812.903],
      "('', 'phil3_22')": [-19812.903, 7737.419, -19812.903],
      "('', 'phil3_33')": [0, 0, 27550.322],
      "('phil3_11', 'phil3_11')": [929672417.073, 1295705157.301, 1295705157.301],
      "('phil3_11', 'phil3_22')": [1390270692.689, 1390270692.689, 2591410314.602],
      "('phil3_11', 'phil3_33')": [0, 0, -1201139621.914],
      "('phil3_22', 'phil3_22')": [1295705157.301, 929672417.073, 1295705157.301],
      "('phil3_22', 'phil3_33')": [0, 0, -1201139621.914],
      "('phil3_33', 'phil3_33')": [0, 0, 835106881.686]
    }
  }
}
\end{minted}

\subsection{Correlated uncertainties}
The two SP mode examples below implement two observables, the branching ratios for $B_s\to\mu^+\mu^-$ and $B^0\to\mu^+\mu^-$, with correlated theoretical uncertainties, computed using \texttt{flavio}. For brevity, we show only the dependence of the observables on the real and imaginary components of a subset of four WET Wilson coefficients defined by the \texttt{"flavio"}  \texttt{WCxf} basis, and information about input parameters in the \texttt{reproducibility} field is omitted. The first example considers the simplified case where only the parameter-independent SM uncertainties are retained, while the second includes parameter-dependent uncertainties.
\vskip-3pt
\textbf{Parameter-independent uncertainties}
\vspace{-4pt}
\begin{minted}[frame=single,framesep=10pt,fontsize=\footnotesize]{json}
{
  "$schema": "https://json.schemastore.org/popxf-1.0.json",
  "metadata": {
    "observable_names": ["BR(Bs->mumu)", "BR(B0->mumu)"],
    "parameters": ["C10_bdmumu", "C10_bsmumu", "C10p_bdmumu", "C10p_bsmumu"],
    "basis": {
      "wcxf": {
        "eft": "WET",
        "basis": "flavio",
        "sectors": ["db", "sb"]
        }
    },
    "scale": 4.8,
    "reproducibility": [{"tool": {"name": "flavio", "version": "2.6.2"}}],
    "misc": {
      "author": ["A. Smolkovic", "P. Stangl"]
    }
  },
  "data": {
    "observable_central": {
      "('', '', 'RR')": [3.629e-09, 1.014e-10],
      "('', 'C10_bdmumu', 'RR')": [0, -4.865e-11],
      "('', 'C10_bsmumu', 'RR')": [-1.742e-09, 0],
      "('', 'C10p_bdmumu', 'RR')": [0, 4.865e-11],
      "('', 'C10p_bsmumu', 'RR')": [1.742e-09, 0],
      "('C10_bdmumu', 'C10_bdmumu', 'II')": [0, 5.838e-12],
      "('C10_bdmumu', 'C10_bdmumu', 'RR')": [0, 5.838e-12],
      "('C10_bdmumu', 'C10p_bdmumu', 'II')": [0, -1.168e-11],
      "('C10_bdmumu', 'C10p_bdmumu', 'RR')": [0, -1.168e-11],
      "('C10_bsmumu', 'C10_bsmumu', 'II')": [1.837e-10, 0],
      "('C10_bsmumu', 'C10_bsmumu', 'RR')": [2.09e-10, 0],
      "('C10_bsmumu', 'C10p_bsmumu', 'II')": [-3.674e-10, 0],
      "('C10_bsmumu', 'C10p_bsmumu', 'RR')": [-4.181e-10, 0],
      "('C10p_bdmumu', 'C10p_bdmumu', 'II')": [0, 5.838e-12],
      "('C10p_bdmumu', 'C10p_bdmumu', 'RR')": [0, 5.838e-12],
      "('C10p_bsmumu', 'C10p_bsmumu', 'II')": [1.837e-10, 0],
      "('C10p_bsmumu', 'C10p_bsmumu', 'RR')": [2.09e-10, 0]
    },
    "observable_uncertainties": {
      "total": [1.046e-10, 5.945e-12]
    }
  }
}
\end{minted}

The parameter-independent \texttt{POPxf} correlation \texttt{JSON} file corresponding to the above observable predictions is given below:
\begin{minted}[frame=single,framesep=10pt,fontsize=\footnotesize]{json}
{
  "$schema": "https://json.schemastore.org/popxf-corr-1.0.json",
  "593771630098eb5325684131f80b4224": {
    "row_names": ["BR(Bs->mumu)", "BR(B0->mumu)"],
    "col_names": ["BR(Bs->mumu)", "BR(B0->mumu)"],
    "correlations": {
      "total":[
        [1.0, 0.407],
        [0.407, 1.0]
      ]
    }
  }
}
\end{minted}

\noindent\textbf{Parameter-dependent uncertainties}

Since $B_s\to\mu^+\mu^-$ and $B^0\to\mu^+\mu^-$ depend on different Wilson coefficients, it is convenient to define them in separate \texttt{POPxf} files. This also simplifies and reduces the size of the corresponding \texttt{POPxf} correlation file. The complete example file containing the full dependence on all the contributing Wilson coefficients and including the entire \texttt{reproducibility} field can be found at \url{https://github.com/pop-xf/examples} under the names \texttt{Bsmumu.json} and \texttt{B0mumu.json}.
\begin{minted}[frame=single,framesep=10pt,fontsize=\footnotesize]{json}
{
  "$schema": "https://json.schemastore.org/popxf-1.0.json",
  "metadata": {
    "observable_names": ["BR(Bs->mumu)"],
    "parameters": ["C10_bsmumu", "C10p_bsmumu"],
    "basis": {
      "wcxf": {
        "eft": "WET",
        "basis": "flavio",
        "sectors": ["sb"]
      }
    },
    "scale": 4.8,
    "reproducibility": [{"tool": {"name": "flavio", "version": "2.6.2"}}],
    "misc": {
      "author": ["A. Smolkovic", "P. Stangl"]
    }
  },
  "data": {
    "observable_central": {
      "('', '', 'RR')": [3.629e-09],
      "('', 'C10_bsmumu', 'RR')": [-1.742e-09],
      "('', 'C10p_bsmumu', 'RR')": [1.742e-09],
      "('C10_bsmumu', 'C10_bsmumu', 'II')": [1.837e-10],
      "('C10_bsmumu', 'C10_bsmumu', 'RR')": [2.09e-10],
      "('C10_bsmumu', 'C10p_bsmumu', 'II')": [-3.674e-10],
      "('C10_bsmumu', 'C10p_bsmumu', 'RR')": [-4.181e-10],
      "('C10p_bsmumu', 'C10p_bsmumu', 'II')": [1.837e-10],
      "('C10p_bsmumu', 'C10p_bsmumu', 'RR')": [2.09e-10]
    },
    "observable_uncertainties": {
      "total": {
        "('', '', 'RR')": [1.046e-10],
        "('', 'C10_bsmumu', 'RR')": [4.653e-11],
        "('', 'C10p_bsmumu', 'RR')": [4.653e-11],
        "('C10_bsmumu', 'C10_bsmumu', 'II')": [4.758e-12],
        "('C10_bsmumu', 'C10_bsmumu', 'RR')": [5.427e-12],
        "('C10_bsmumu', 'C10p_bsmumu', 'II')": [9.516e-12],
        "('C10_bsmumu', 'C10p_bsmumu', 'RR')": [1.085e-11],
        "('C10p_bsmumu', 'C10p_bsmumu', 'II')": [4.758e-12],
        "('C10p_bsmumu', 'C10p_bsmumu', 'RR')": [5.427e-12]
      }
    }
  }
}
\end{minted}
\begin{minted}[frame=single,framesep=10pt,fontsize=\footnotesize]{json}
{
  "$schema": "https://json.schemastore.org/popxf-1.0.json",
  "metadata": {
    "observable_names": ["BR(B0->mumu)"],
    "parameters": ["C10_bdmumu", "C10p_bdmumu"],
    "basis": {
      "wcxf": {
        "eft": "WET",
        "basis": "flavio",
        "sectors": ["db"]
      }
    },
    "scale": 4.8,
    "reproducibility": [{"tool": {"name": "flavio", "version": "2.6.2"}}],
    "misc": {
      "author": ["A. Smolkovic", "P. Stangl"]
    }
  },
  "data": {
    "observable_central": {
      "('', '', 'RR')": [1.014e-10],
      "('', 'C10_bdmumu', 'RR')": [-4.865e-11],
      "('', 'C10p_bdmumu', 'RR')": [4.865e-11],
      "('C10_bdmumu', 'C10_bdmumu', 'II')": [5.838e-12],
      "('C10_bdmumu', 'C10_bdmumu', 'RR')": [5.838e-12],
      "('C10_bdmumu', 'C10p_bdmumu', 'II')": [-1.168e-11],
      "('C10_bdmumu', 'C10p_bdmumu', 'RR')": [-1.168e-11],
      "('C10p_bdmumu', 'C10p_bdmumu', 'II')": [5.838e-12],
      "('C10p_bdmumu', 'C10p_bdmumu', 'RR')": [5.838e-12]
    },
    "observable_uncertainties": {
      "total": {
        "('', '', 'RR')": [5.945e-12],
        "('', 'C10_bdmumu', 'RR')": [2.805e-12],
        "('', 'C10p_bdmumu', 'RR')": [2.805e-12],
        "('C10_bdmumu', 'C10_bdmumu', 'II')": [3.345e-13],
        "('C10_bdmumu', 'C10_bdmumu', 'RR')": [3.345e-13],
        "('C10_bdmumu', 'C10p_bdmumu', 'II')": [6.691e-13],
        "('C10_bdmumu', 'C10p_bdmumu', 'RR')": [6.691e-13],
        "('C10p_bdmumu', 'C10p_bdmumu', 'II')": [3.345e-13],
        "('C10p_bdmumu', 'C10p_bdmumu', 'RR')": [3.345e-13]
      }
    }
  }
}
\end{minted}

The corresponding \texttt{POPxf} correlation \texttt{JSON} file is given below. The $(0,0,0,0)$ entry of each numerical dataset corresponds to the parameter-independent piece documented in the example above. The remaining terms account for the parameter dependence. Since each data file describes one observable and has nine entries in \texttt{observable\_central}, each numerical dataset is represented by an array of shape $(1,1,9,9)$. The complete correlation file containing the full dependence on all the contributing Wilson coefficients can be found at \url{https://github.com/pop-xf/examples} under the name \texttt{corr.json}.

\begin{minted}[frame=single,framesep=10pt,fontsize=\footnotesize]{json}
{
  "$schema": "https://json.schemastore.org/popxf-corr-1.0.json",
  "974bcd243772ce08f33a16c7fda240de": {
    "row_names": ["BR(B0->mumu)"],
    "col_names": ["BR(B0->mumu)"],
    "correlations": {
      "total": [[[
            [1.0, -0.994, 0.994, 0.977, 0.977, -0.977, -0.977, 0.977, 0.977],
            [-0.994, 1.0, -1.0, -0.994, -0.994, 0.994, 0.994, -0.994, -0.994],
            [0.994, -1.0, 1.0, 0.994, 0.994, -0.994, -0.994, 0.994, 0.994],
            [0.977, -0.994, 0.994, 1.0, 1.0, -1.0, -1.0, 1.0, 1.0],
            [0.977, -0.994, 0.994, 1.0, 1.0, -1.0, -1.0, 1.0, 1.0],
            [-0.977, 0.994, -0.994, -1.0, -1.0, 1.0, 1.0, -1.0, -1.0],
            [-0.977, 0.994, -0.994, -1.0, -1.0, 1.0, 1.0, -1.0, -1.0],
            [0.977, -0.994, 0.994, 1.0, 1.0, -1.0, -1.0, 1.0, 1.0],
            [0.977, -0.994, 0.994, 1.0, 1.0, -1.0, -1.0, 1.0, 1.0]
          ]]]
    }
  },
  "a262ca783a3dd055c77ec5c6c75c6ffe": {
    "row_names": ["BR(B0->mumu)"],
    "col_names": ["BR(Bs->mumu)"],
    "correlations": {
      "total": [[[
            [0.407, -0.389, 0.389, 0.35, 0.349, -0.35, -0.349, 0.35, 0.349],
            [-0.367, 0.371, -0.371, -0.356, -0.355, 0.356, 0.355, -0.356, -0.355],
            [0.367, -0.371, 0.371, 0.356, 0.355, -0.356, -0.355, 0.356, 0.355],
            [0.322, -0.347, 0.347, 0.358, 0.357, -0.358, -0.357, 0.358, 0.357],
            [0.322, -0.347, 0.347, 0.358, 0.357, -0.358, -0.357, 0.358, 0.357],
            [-0.322, 0.347, -0.347, -0.358, -0.357, 0.358, 0.357, -0.358, -0.357],
            [-0.322, 0.347, -0.347, -0.358, -0.357, 0.358, 0.357, -0.358, -0.357],
            [0.322, -0.347, 0.347, 0.358, 0.357, -0.358, -0.357, 0.358, 0.357],
            [0.322, -0.347, 0.347, 0.358, 0.357, -0.358, -0.357, 0.358, 0.357]
          ]]]
    }
  },
  "1af389d015582d6903a33587d94d45ea": {
    "row_names": ["BR(Bs->mumu)"],
    "col_names": ["BR(Bs->mumu)"],
    "correlations": {
      "total": [[[
            [1.0, -0.978, 0.978, 0.848, 0.902, -0.848, -0.902, 0.848, 0.902],
            [-0.978, 1.0, -1.0, -0.914, -0.972, 0.914, 0.972, -0.914, -0.972],
            [0.978, -1.0, 1.0, 0.914, 0.972, -0.914, -0.972, 0.914, 0.972],
            [0.848, -0.914, 0.914, 1.0, 0.939, -1.0, -0.939, 1.0, 0.939],
            [0.902, -0.972, 0.972, 0.939, 1.0, -0.939, -1.0, 0.939, 1.0],
            [-0.848, 0.914, -0.914, -1.0, -0.939, 1.0, 0.939, -1.0, -0.939],
            [-0.902, 0.972, -0.972, -0.939, -1.0, 0.939, 1.0, -0.939, -1.0],
            [0.848, -0.914, 0.914, 1.0, 0.939, -1.0, -0.939, 1.0, 0.939],
            [0.902, -0.972, 0.972, 0.939, 1.0, -0.939, -1.0, 0.939, 1.0]
          ]]]
    }
  }
}
\end{minted}

%%%%%%%%%%%%%%%%%%%%%%%%%%%%%%%%%%%%%%%%%%%%%%%
\bibliographystyle{JHEP}
\bibliography{biblio}

\end{document}